\let\LaTeXlabel\label \documentclass[aps,physrev,reprint,superscriptaddress,amsmath,amssymb,floatfix,longbibliography]{revtex4-2}
\usepackage{graphicx}
\usepackage{bm}
\usepackage{upgreek}   \usepackage{xcolor}
\makeatletter\let\label\LaTeXlabel\let\ltx@label\label\makeatother
\usepackage[colorlinks=true,linkcolor=blue!45!black,citecolor=blue!45!black,urlcolor=blue!45!black]{hyperref}

\newif\ifplaceholder
\placeholderfalse
\newif\ifsimulated
\simulatedfalse
\newif\ifmatched
\matchedtrue
\newif\ifcharknown
\charknowntrue
\newif\ifRecMatched
\RecMatchedtrue
\newif\ifRecActMatched
\RecActMatchedtrue
\newif\ifPasGfAllIn
\PasGfAllIntrue
\newif\ifActAllIn
\ActAllIntrue
\newif\ifPasLowBelow
\PasLowBelowfalse
\newif\ifPasHaveGf
\PasHaveGftrue
\newif\ifPasSrcAll
\PasSrcAlltrue
\newif\ifPasMeasCurves
\PasMeasCurvestrue
\newif\ifActTgtBelow
\ActTgtBelowtrue
\newif\ifActFieldHiNeg
\ActFieldHiNegtrue
\newif\ifActBgHiNeg
\ActBgHiNegfalse
\newif\ifActMeasCurves
\ActMeasCurvestrue
\newif\ifMeanSceneSqBright
\MeanSceneSqBrighttrue
\newif\ifGfRecords
\GfRecordstrue
\newif\ifSlotTest
\SlotTesttrue
\newif\ifSlotRef
\SlotReftrue
\newif\ifPasFlatRej
\PasFlatRejtrue
\newif\ifPasGfOk
\PasGfOktrue
\newif\ifMeanSigPas
\MeanSigPastrue
\newif\ifMeanSigPasHi
\MeanSigPasHitrue
\newif\ifMeanSigAct
\MeanSigActfalse
\newcommand{\NskyPas}{208}
\newcommand{\NframeAct}{100}
\newcommand{\XlowDb}{-20}
\newcommand{\XhighDb}{+5}
\newcommand{\PsnrCohLow}{15.3}
\newcommand{\PsnrSqLow}{21.2}
\newcommand{\PsnrCohHigh}{47.2}
\newcommand{\PsnrSqHigh}{47.4}
\newcommand{\EsnrLowDb}{-20}
\newcommand{\EsnrHighDb}{-10}
\newcommand{\PsnrActCohLow}{14.3}
\newcommand{\PsnrActSqLow}{19.1}
\newcommand{\PsnrActCohHigh}{27.0}
\newcommand{\PsnrActSqHigh}{30.5}

\newcommand{\gval}{0.495}
\newcommand{\Gdb}{3.05}
\newcommand{\TwoGdb}{6.10}
\newcommand{\xstarDb}{-1.5}
\newcommand{\PfieldMtwenty}{5.9}
\newcommand{\PfieldZero}{0.8}
\newcommand{\PfieldPfive}{0.2}

\newcommand{\XcBg}{-6.3}
\newcommand{\XcSrc}{-11.8}

\newcommand{\ThBgMfive}{2.5}
\newcommand{\TOvC}{20.8 \pm 0.3}
\newcommand{\TOvS}{24.7 \pm 0.4}

\newcommand{\TOvG}{3.95}

\newcommand{\TOvGpm}{0.23}
\newcommand{\TOvP}{4.05}

\newcommand{\TTgC}{16.2 \pm 0.4}
\newcommand{\TTgS}{19.4 \pm 0.5}

\newcommand{\TTgG}{3.15}

\newcommand{\TTgGpm}{0.31}
\newcommand{\TTgP}{3.32}

\newcommand{\TBgC}{24.3 \pm 0.3}
\newcommand{\TBgS}{30.2 \pm 0.3}

\newcommand{\TBgDshort}{5.9}
\newcommand{\TBgG}{5.94}

\newcommand{\TBgGpm}{0.23}
\newcommand{\TBgP}{5.92}

\newcommand{\RhoTgt}{4.6}

\newcommand{\GrecDb}{3.05}

\newcommand{\NrecPairs}{102}

\newcommand{\ScopeClearDb}{14.8}

\newcommand{\CohExcessDb}{0.01}

\newcommand{\TafterRms}{1.9}

\newcommand{\XspanMin}{19.0}
\newcommand{\XspanMax}{24.3}

\newcommand{\CohExcessErr}{0.03}

\newcommand{\ActBgMin}{5.94}
\newcommand{\ActBgMax}{6.35}
\newcommand{\TqMeas}{3.7\times10^{10}}
\newcommand{\GfeLo}{34}
\newcommand{\GfeHi}{60}
\newcommand{\VpiImplied}{267}

\newcommand{\RecXzero}{-9.6}

\newcommand{\RecActSnr}{-14.8}

\newcommand{\RecSigMax}{0.11}

\newcommand{\PasResHiMax}{0.08}
\newcommand{\PasResLoA}{0.13}
\newcommand{\PasResLoB}{0.15}
\newcommand{\PasCiLoA}{0.13}
\newcommand{\PasCiLoB}{0.07}
\newcommand{\PasResGfMax}{0.06}

\newcommand{\PasSrcResMax}{0.30}

\newcommand{\PasLineDevMax}{0.07}

\newcommand{\GeffPasMin}{3.01}
\newcommand{\GeffPasMax}{3.08}
\newcommand{\GfLoDb}{3.14}
\newcommand{\GfHiDb}{2.85}
\newcommand{\XcBgMeas}{-5.6}
\newcommand{\XcSrcMeas}{-11.1}
\newcommand{\XcOffMeas}{5.5}

\newcommand{\SrcRatioHiMax}{0.6}
\newcommand{\ActBgResMax}{0.09}

\newcommand{\WanderActSqPowMax}{9.1}

\newcommand{\WanderActCohPowMax}{2.4}
\newcommand{\GeffActMin}{2.97}
\newcommand{\GeffActMax}{3.14}

\newcommand{\ActBlock}{20}

\newcommand{\ActNblocks}{5}

\newcommand{\GeffLow}{3.14}

\newcommand{\ActFieldResMax}{0.10}

\newcommand{\ActLineDevMax}{0.18}

\newcommand{\FBResAllMax}{0.15}
\newcommand{\ActTgtResMax}{0.17}

\newcommand{\MeanMseFrac}{0.2}

\newcommand{\SlotRatioCoh}{1.030}
\newcommand{\SlotRatioSq}{1.005}
\newcommand{\SlotRatioCohClean}{0.997}
\newcommand{\SlotRatioSqClean}{1.000}
\newcommand{\SlotErr}{0.002}
\newcommand{\SlotLinesCoh}{5.5 and 17.0}

\newcommand{\ActTgtFirst}{6.08}
\newcommand{\ActTgtLast}{3.06}
\newcommand{\ActWanderEffMax}{0.25}

\newcommand{\GeffTable}{2.97}
\newcommand{\TwoGeffTable}{5.94}
\definecolor{phcol}{HTML}{B4530D}
\definecolor{phbg}{HTML}{FFF5EA}
\newcommand{\ph}[1]{\ifplaceholder\textcolor{phcol}{#1}\else#1\fi}

\newcommand{\Var}{\operatorname{Var}}
\newcommand{\E}{\mathbb{E}}
\newcommand{\PSNR}{\mathrm{PSNR}}
\newcommand{\dPSNR}{\Delta\mathrm{PSNR}}
\newcommand{\Ihat}{\hat I}
\newcommand{\Vhat}{\hat V}

\begin{document}
\raggedbottom

\title{Quantum-Enhanced Passive and Active\texorpdfstring{\\}{} Radio-Frequency Photonic Imaging}

\author{Visuttha Manthamkarn}
\affiliation{Department of Electrical Engineering and Computer Science, University of Michigan, Ann Arbor, Michigan 48109, USA}
\author{Haowei Shi}
\affiliation{Ming Hsieh Department of Electrical and Computer Engineering, University of Southern California, Los Angeles, California 90089, USA}
\author{Shuaifeng Guo}
\email[Contact author: ]{shuaiguo@umich.edu}
\affiliation{Department of Electrical Engineering and Computer Science, University of Michigan, Ann Arbor, Michigan 48109, USA}
\author{Christopher M. Jones}
\affiliation{Halliburton Technology, 3000 N Sam Houston Pkwy E, Houston, Texas 77032, USA}
\author{Quntao Zhuang}
\affiliation{Ming Hsieh Department of Electrical and Computer Engineering, University of Southern California, Los Angeles, California 90089, USA}
\author{Zheshen Zhang}
\email[Contact author: ]{zszh@umich.edu}
\affiliation{Department of Electrical Engineering and Computer Science, University of Michigan, Ann Arbor, Michigan 48109, USA}

\date{\today}

\begin{abstract}
Radio-frequency (RF) photonics enables broadband, phase-coherent sensing beyond the reach of purely electronic architectures, yet at a given optical carrier power its sensitivity is bounded by the standard quantum limit of the optical readout. Here, we experimentally demonstrate that this limit can be surpassed in both passive and active RF imaging modalities by encoding RF signals onto a bright phase-squeezed optical carrier and recovering them via balanced homodyne detection, while the RF front end, illumination, and image reconstruction remain classical. A unified measurement model shows that phase squeezing reduces the variance of the additive readout noise by a single factor $g$, yielding peak signal-to-noise ratio (PSNR) gains of up to $2G$, where $G\equiv10\log_{10}(1/g)$: between $G$ and $2G$ for deterministic targets in active imaging when the classical noise of the echoes is negligible against the readout noise, and set by the ratio of scene noise to readout noise in passive imaging. We validate the model in an experiment in the 5--20~MHz band, in which synthesized RF scenes are replayed through a physical squeezed-light receiver. With \ph{3}~dB of squeezing, the PSNR improves by about \ph{6}~dB, close to $2G$, in the readout-noise-dominated regime of both modalities, and the model predicts the measured gains without free parameters. These results identify where squeezed-light readout benefits RF-photonic imaging: wherever the optical readout noise, rather than the scene or the RF front end, dominates the image error.
\end{abstract}

\maketitle

\ifplaceholder
\ifsimulated
\noindent\fcolorbox{phcol}{phbg}{\parbox{\dimexpr\columnwidth-2\fboxsep-2\fboxrule\relax}{\footnotesize\textbf{Draft with simulated results.} Figures~\ref{fig:passive}--\ref{fig:record}, Table~\ref{tab:active} and all values printed in \textcolor{phcol}{this color} come in part from a digital twin of the 5--20~MHz measurement (same parameters, $g=\gval$) and stand in for the measured data; the statements about measured data, also in this color, are to be checked when the data replace them.}}
\else
\noindent\fcolorbox{phcol}{phbg}{\parbox{\dimexpr\columnwidth-2\fboxsep-2\fboxrule\relax}{\footnotesize\textbf{Draft with measured results.} Figures~\ref{fig:passive}--\ref{fig:record}, Table~\ref{tab:active} and the values printed in \textcolor{phcol}{this color} are generated from the measurement by the analysis; values and statements in this color are to be checked by the authors before submission.}}
\fi
\fi

\section{Introduction}
Radio-frequency (RF) sensing and imaging underpin a wide range of applications, including remote sensing~\cite{Ulaby1981,Moreira2013}, radar~\cite{Skolnik2001,Cumming2005}, navigation~\cite{Misra2006,Kaplan2017}, and spectrum monitoring~\cite{Yucek2009,Haykin2005}. Distributed RF-photonic sensor networks and large-scale interferometric arrays have emerged as particularly promising platforms for passive and active imaging and surveillance~\cite{Hotan2021,Dong2020}. In recent years, RF-photonic architectures have become a powerful platform for such systems, enabling broadband operation, low-loss signal distribution, and precise phase-coherent processing beyond the limits of purely electronic approaches~\cite{Dong2020,Seeds2006,Ghelfi2014,Pan2020,Wang2020,Marpaung2019}. These continued advantages drive RF-photonic receivers to quickly approach the sensitivity limit set by measurement noise at the optical readout stage, which at a given optical power is the standard quantum limit (SQL) of optical readout~\cite{Caves1981,Yuen1983,Haus2004}---with shot-noise-limited operation demonstrated in RF-photonic links~\cite{Williams1996,Yao2023} and in the optical readout of an optoelectromechanical RF transducer~\cite{Bagci2014}.

Quantum resources offer a potential route to overcoming this limitation. Squeezed states of light have been widely explored in precision optical metrology, where engineering the quantum fluctuations of the optical field below the shot-noise level enables sensitivity beyond the SQL at a fixed optical power level~\cite{Giovannetti2011,Lawrie2019}. This principle has been most prominently demonstrated in gravitational-wave interferometers, where squeezed-light readout has led to substantial improvements in detection sensitivity~\cite{LIGO2011,Aasi2013,Tse2019,Acernese2019,Lough2021,Ganapathy2023}. These successes raise the question of whether similar quantum noise suppression can be harnessed in RF-photonic imaging systems.

Previous efforts toward quantum-enhanced RF sensing have largely focused on nonclassical illumination and quantum receivers, known as quantum illumination protocols~\cite{Tan2008,Barzanjeh2015,Assouly2023,Zhang2015}. While powerful in principle, such approaches often require entanglement distribution, quantum storage, low probe power, and joint quantum measurements of the returned and retained modes, posing significant challenges for practical implementation in RF systems~\cite{Shapiro2020}. More recently, quantum enhancement via squeezed optical readout has been explored in specific RF-photonic measurement tasks, including angle-of-arrival estimation and distributed RF amplitude measurements~\cite{Xia2020,Li2023,Li2024}. These experiments characterize the quantum advantage at the level of RF-signal estimation and resolution. In imaging, the advantage must instead survive the statistical estimation and reconstruction that form an image from an ensemble of measurements---second order in the measured fields for interferometry---and the image error also contains fluctuations of the scene itself, which squeezing cannot reduce. Shi \textit{et al.}~\cite{Shi2025} proposed a squeezed-readout RF-photonic imaging protocol and considered a single active synthetic-aperture radar modality in a numerical study, demonstrating that squeezing can overcome the noise penalty of low transduction efficiency in subterranean synthetic-aperture sensing. What remains open is how a single, modality-independent reduction of the readout noise carries through the different statistical estimators that form passive interferometric and active synthetic-aperture images, and how much of it reaches each image.

Here, we present a quantum-enhanced RF-photonic imaging architecture that introduces quantum noise suppression exclusively at the optical readout stage at the receiver. The key idea is to encode RF information onto a bright phase-squeezed optical signal and to recover this information using phase-sensitive homodyne detection. By reducing phase-quadrature fluctuations of the optical signal at the receiver, the architecture suppresses additive measurement noise below the SQL while leaving the RF front end, illumination waveforms, and digital reconstruction pipeline entirely classical. This establishes a practical pathway toward quantum-enhanced RF-photonic imaging that is compatible with existing RF-photonic systems whose sensitivity is limited by optical readout noise and scalable to more complex sensing and imaging architectures.

To quantify the advantage enabled by our RF-photonic imaging architecture, we develop a unified measurement model that shows how phase squeezing produces a predictable rescaling of measurement noise, independent of sensing modality. This model applies equally to passive interferometric imaging, which infers spatial information from naturally emitted RF radiation, and to active synthetic-aperture imaging, which relies on coherent processing of transmitted and reflected waveforms. The reduction of the readout noise is the same in both modalities, but the image gain is not, because the estimators that form the images weight the readout noise differently. For a readout-noise variance reduced by the factor $g$, the gain lies between 0 and $2G$ in passive imaging, where $G\equiv10\log_{10}(1/g)$, depending on the ratio of the scene noise to the readout noise, and between $G$ and $2G$ for a deterministic target in active imaging whose echoes carry negligible classical noise. We experimentally validate this architecture in the 5--20~MHz high-frequency (HF) band, where the RF waveform drives the electro-optic modulator directly, without frequency conversion, in an experiment in which a single physical receiver is time-division multiplexed across all antenna and aperture channels. A preliminary demonstration with an earlier implementation of the experiment was presented in Ref.~\cite{Manthamkarn2026}. We demonstrate improvements in image quality, quantified by the peak signal-to-noise ratio (PSNR), in both modalities, together with their dependence on the ratio of scene noise to readout noise. With \ph{\GrecDb}~dB of squeezing in the detected records, the PSNR improves by up to \ph{\PfieldMtwenty}~dB in passive imaging where the readout noise dominates and by \ph{\TBgDshort}~dB on the background of active images at an echo signal-to-noise ratio (SNR) of $-15$~dB, close to the limit $2G=\TwoGdb$~dB. \ph{The model predicts the measured gains without free parameters.}

\section{Quantum-Enhanced RF-Photonic Receiver Architecture}

\begin{figure*}[t]
\includegraphics[width=\textwidth]{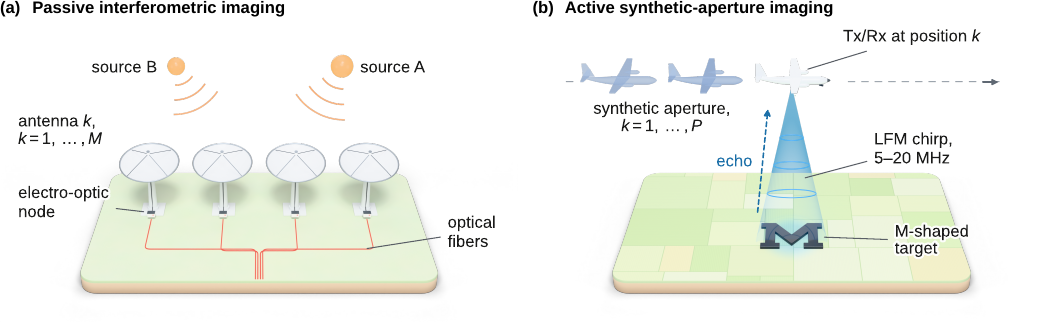}\\[5pt]
\includegraphics[width=\textwidth]{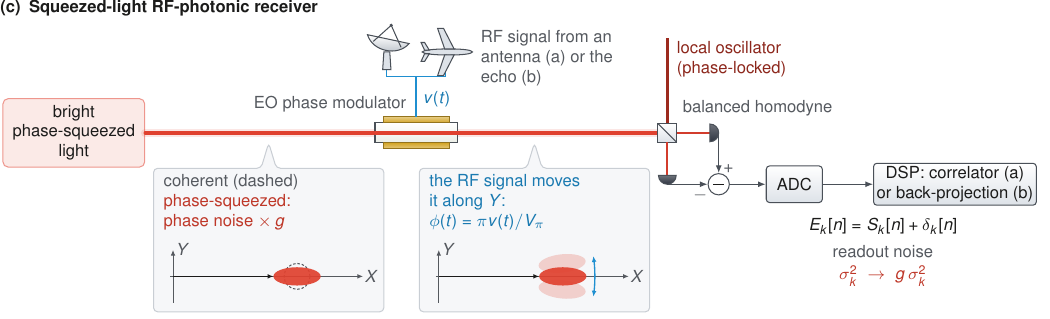}
\caption{Quantum-enhanced RF-photonic imaging. (a)~Passive interferometric imaging: antennas, each with an electro-optic node, send their signals over optical fibers to squeezed-light receivers and a correlator, which forms the visibilities $\Vhat_{ij}$ between receiver pairs and synthesizes the image $\Ihat(l,m)$ of the radio sky ($l,m$: direction cosines). (b)~Active synthetic-aperture imaging: a platform with a transmitter and receiver (Tx/Rx) sends linear frequency-modulated (LFM) chirps at aperture positions $k=1,\dots,P$ along its flight path and records the echoes $E_k[n]$; back-projection gives the focused field $z_q$ at pixel $q$ and the image $I_q=|z_q|^2$. The scenes are illustrations; in the experiment a single receiver records the replayed channels one after another (Appendix~\ref{app:channels}). (c)~Squeezed-light RF-photonic receiver: the RF voltage $v(t)$ of an antenna in (a) or of the echo in (b) drives an electro-optic (EO) phase modulator with half-wave voltage $V_\pi$, which shifts the phase of a bright phase-squeezed optical carrier by $\phi(t)=\pi v(t)/V_\pi$; balanced homodyne detection against a local oscillator, locked to the phase quadrature, and an analog-to-digital converter (ADC) give the record $E_k[n]=S_k[n]+\delta_k[n]$ of channel $k$, the RF signal plus the readout noise [Eq.~\eqref{eq:Ek}], which digital signal processing (DSP) correlates (a) or back-projects (b). Below the beam: the carrier in phase space (quadratures $X$ and $Y$) before and after the modulator. Its phase-squeezed noise (red ellipse) has the variance of a coherent state's (dashed circle) times $g$ along the phase quadrature $Y$; the phase shift $\phi(t)$ moves it along $Y$ (faint ellipses, exaggerated), the quadrature that the homodyne detection reads.}
\label{fig:architecture}
\end{figure*}

Figure~\ref{fig:architecture} illustrates a unified architecture for both passive [Fig.~\ref{fig:architecture}(a)] and active [Fig.~\ref{fig:architecture}(b)] RF-photonic imaging, despite their fundamental differences in signal origin and reconstruction. In passive imaging, RF signals naturally emitted or scattered by the scene are collected across a distributed antenna array and coherently transduced to the optical domain at each node without invoking any active transmitter; spatial information is subsequently recovered from cross-correlations between antenna pairs via aperture synthesis. In the active regime, a classical transmitter emits a tailored coherent waveform that illuminates the scene, and backscattered echoes are coherently recorded across multiple aperture positions to synthesize a large effective aperture. Both modalities exploit an identical quantum-enhanced receiver [Fig.~\ref{fig:architecture}(c)]: the received RF waveform drives an electro-optic phase modulator that encodes the RF information onto the phase quadrature of a bright optical carrier held in a phase-squeezed state. The phase-modulated field is then measured by balanced homodyne detection, converting the signal back into the RF domain for post-processing. The complete experimental implementation is detailed in Appendix~\ref{app:setup}, covering the optical source and squeezed-state generation (Appendix~\ref{app:source}), electro-optic transduction and homodyne detection (Appendix~\ref{app:readout}), the synthesized RF scenes (Appendix~\ref{app:scenes}), and RF channel implementation and signal processing (Appendix~\ref{app:channels}).

To understand how quantum noise suppression is achieved within this architecture, we describe the optical state preparation and detection protocol employed at the receiver. Engineering quantum fluctuations of the optical field at the detector is a well-established strategy in precision optical metrology~\cite{Giovannetti2011,Lawrie2019,Xiao1987,Atkinson2021}. Here, this paradigm is extended to RF-photonic imaging by preparing the optical carrier in a bright phase-squeezed state whose reduced-noise quadrature is aligned with the RF-induced phase modulation. Balanced homodyne detection measures this phase quadrature with sensitivity limited by optical shot noise for a coherent carrier~\cite{Caves1981,Yuen1983,Haus2004}. Phase squeezing reduces the variance of the measured quadrature below the shot noise, thereby suppressing additive measurement noise at the homodyne output while leaving the encoded RF signal unchanged. Because this noise reduction acts solely on the optical measurement process, it propagates deterministically through all subsequent RF processing stages, providing a modality-independent rescaling of the additive readout noise governed entirely by a single parameter, the squeezing factor $g$ of Sec.~\ref{sec:model}.

Confining the quantum resource to the optical readout, while the RF front end, propagation channel, illumination waveform, and digital reconstruction remain classical, provides two complementary advantages. First, it suppresses additive measurement noise below the SQL of optical phase measurement without nonclassical RF illumination, which is inherently limited by low probe power~\cite{Tan2008,Assouly2023,Shapiro2020}. Second, it allows the architecture to be integrated into established RF-photonic radar and imaging pipelines with minimal modification to transmitters, antenna arrays, or digital back-end processors~\cite{Dong2020,Ghelfi2014,Pan2020,Wang2020}.

\section{Quantum-Enhanced RF-Photonic Receiver Performance}

The quantum-enhanced performance of the receiver architecture can be understood within a unified measurement model that applies equally to passive and active RF sensing. We experimentally validate this model for both modalities using the same optical readout system, operated with either a coherent or a bright phase-squeezed carrier of nominally equal power and compared at equal RF signal (Sec.~\ref{sec:receiver}). The RF scenes are synthesized and replayed through the physical receiver by an arbitrary waveform generator (AWG), and the images are formed from the digitized homodyne records, so the readout noise in the images is that of the receiver. The experiment therefore tests the receiver and the image formation, while the antennas, RF propagation, or platform motion of a deployed system are modeled and emulated (Sec.~\ref{sec:discussion}). Unless stated otherwise, uncertainties are 95\% confidence intervals.

\subsection{Unified measurement model}\label{sec:model}
Following optical detection, the directly sampled RF record of each measurement channel $k$---denoting the receiver index in passive imaging or the pulse and aperture-position index in active imaging---is converted to discrete-time complex samples (frequency channels in passive imaging, the analytic signal in active imaging),
\begin{equation}
  E_k[n] = S_k[n] + \delta_k[n],
  \label{eq:Ek}
\end{equation}
where $S_k[n]$ denotes the RF-dependent signal component (a thermal field in passive imaging, a deterministic echo in active imaging) and $\delta_k[n]$ represents the readout noise, the additive measurement noise arising from optical shot noise and electronic noise.

For a coherent carrier, the variance of $\delta_k[n]$ is the SQL of optical phase measurement at the power of the optical carrier, plus the electronic noise of the detection:
\begin{equation}
  \E\bigl[|\delta_k[n]|^2\bigr] = \sigma_k^2 .
  \label{eq:coh}
\end{equation}
The receiver-dependent noise variance $\sigma_k^2$ captures heterogeneity across channels, such as unequal optical powers or detector efficiencies in a distributed receiver array; in active imaging, it is the variance per sample of an equivalent white noise (Appendix~\ref{app:active_rc}).

When the optical field is prepared in a bright phase-squeezed state aligned to the RF-modulated phase quadrature, the additive noise variance is reduced to
\begin{equation}
  \E\bigl[|\delta_k[n]|^2\bigr] = g\,\sigma_k^2, \qquad 0<g<1,
  \label{eq:sq}
\end{equation}
where $g\equiv\sigma_{\rm sq}^2/\sigma_{\rm coh}^2$ is the ratio of the homodyne noise variances with the phase-squeezed and the coherent carrier, while $S_k[n]$ remains unchanged. Because the electronic noise of the detection is not squeezed, $g$ describes the total readout noise rather than the optical state alone: for a phase-quadrature variance $s$ of the squeezed light, a detection efficiency $\eta$ and an electronic noise $\varepsilon$, with $s$ and $\varepsilon$ relative to the shot noise, $g=(\eta s+1-\eta+\varepsilon)/(1+\varepsilon)$. In the experiment $g$ is measured from the records themselves (Sec.~\ref{sec:receiver}). Quantum enhancement therefore acts exclusively through suppression of the additive readout noise, with the squeezing factor $g$ determined solely by the optical readout stage and independent of sensing modality. We assume a common squeezing factor $g$ across all receivers, corresponding to identically configured readout stages, each with its own squeezed carrier; generalization to receiver-dependent squeezing $g_k$ is straightforward.

How much of this reduction reaches the image depends on the noise already carried by the RF signal. We characterize it by the operating point $x\equiv P_k/\sigma_k^2$, the ratio of the RF signal power $P_k\equiv\E\bigl[|S_k[n]|^2\bigr]$ of passive imaging, itself thermal noise, to the coherent readout noise; in active imaging, $x$ is the same ratio for the classical noise in the echoes, not for the echoes themselves (Sec.~\ref{sec:active}). Referred to the modulator input of impedance $R$, shot-noise-limited readout of a carrier of power $P_{\rm opt}$ and optical frequency $\nu$ is equivalent to the thermal noise of a matched resistor at the temperature $T_{\rm q}=(V_\pi/\pi)^2\,h\nu/(2\eta P_{\rm opt}k_BR)$, where $V_\pi$ is the half-wave voltage of the modulator ($\phi=\pi v/V_\pi$) and both noises are one-sided voltage spectral densities across the input (Appendix~\ref{app:x0}). A passive receiver therefore operates at $x=(G_{\rm fe}T_A+T_N)/T_{\rm q}$, where $T_A$ is the antenna temperature, $G_{\rm fe}$ the gain of any RF front end ($G_{\rm fe}=1$ without one), and $T_N$ the front end's own noise at its output, set by its noise figure. Referred to the antenna, this noise, $T_N/G_{\rm fe}$, is negligible against the HF sky, so that $x=G_{\rm fe}T_A/T_{\rm q}$ here; it is not negligible where the scene is cold, as in superconducting-qubit readout or sensing at higher frequencies. With coherent readout, $x$ is the fractional rise of the homodyne noise power when the RF signal is applied; we quote it at the 12.5-MHz band center as $x_0$. Squeezing lowers $T_{\rm q}$ by the factor $g$, as would a carrier power raised to $P_{\rm opt}/g$; it therefore matters where the carrier power cannot be raised, for example because of the optical power budget of a distributed array, the power handling of the modulator and the detectors, nonlinear effects, or measurement-induced backactions.

Because all subsequent RF imaging operations---visibility estimation, matched filtering, interpolation, and coherent summation---are linear (active imaging) or bilinear (passive imaging) in $E_k[n]$, this multiplicative noise reduction propagates deterministically through the imaging pipeline (Appendices~\ref{app:passive} and~\ref{app:active}).

\subsection{Readout noise and operating points}\label{sec:receiver}

Both experiments use the same receiver (Appendix~\ref{app:setup}). Without an RF signal, the homodyne noise of the coherent carrier exceeds the shot noise of the local oscillator by only \ph{$\CohExcessDb\pm\CohExcessErr$}~dB over the 5--20~MHz band, indicating that the coherent readout operates at the SQL. The phase-squeezed carrier at the same power lowers the readout noise in the records by $G=\ph{\GrecDb}$~dB ($g=\gval$), averaged over the \ph{\NrecPairs} recording pairs of both experiments (a recording: the records of one data set at one operating point or echo SNR), and leaves the RF signal unchanged (Appendix~\ref{app:readout}); the model curves below use this $g$. The squeezing weakens toward the top of the band.

The operating points are measured from the same records by projecting each record onto the known replayed waveform (Appendix~\ref{app:readout}). In passive imaging, the operating point $x(f)$ at RF frequency $f$ falls by \ph{\XspanMin}--\ph{\XspanMax}~dB from 5 to 20~MHz, mostly because the brightness of the HF sky falls steeply with frequency~\cite{Cane1979}; $x_0$ therefore only labels an operating point, and predictions below use the whole measured $x(f)$. Each readout's images are referred to its own measured signal, and the predictions use the squeezing level $G_{\rm eff}$ of each recording referred to its signal, \ph{\GeffPasMin}--\ph{\GeffPasMax}~dB on average at each passive operating point and \ph{\GeffActMin}--\ph{\GeffActMax}~dB in active imaging, without a trend with the operating point (Appendices~\ref{app:readout} and~\ref{app:channels}). Every prediction thus uses only quantities measured at the receiver or known from the replayed scenes; none is adjusted to the measured gains.

\subsection{PSNR gain in passive interferometric imaging}\label{sec:passive}

In passive interferometric imaging~\cite{Thompson2017}, an array of $M$ receivers collects RF signals $S_k[n]$ originating from thermal radiation emitted by the scene, where $k\in\{1,\dots,M\}$ indexes individual receivers. The noise variance $\sigma_k^2$ may differ across receivers in a distributed array. Spatial information is inferred from the complex visibilities between receiver pairs $(i,j)$, estimated in each frequency channel from $N$ samples as
\begin{equation}
  \Vhat_{ij} = \frac{1}{N}\sum_{n} E_i[n]\,E_j^*[n],
  \label{eq:vis}
\end{equation}
which sample the spatial coherence of the scene brightness via the van Cittert--Zernike theorem~\cite{BornWolf1999}. The synthesized real-valued image is then obtained through a weighted inverse Fourier sum over the measured baselines,
\begin{equation}
  \Ihat(l,m) = 2\,\mathrm{Re}\Bigl\{\sum_{b=(i,j)} w_b\,\Vhat_b\,e^{i2\pi(u_bl+v_bm)}\Bigr\},
  \label{eq:image}
\end{equation}
where $(u_b,v_b)$ are the spatial-frequency coordinates of baseline $b$, $(l,m)$ are direction cosines on the sky, and $w_b$ are synthesis weights; the broadband image is the sum of Eq.~\eqref{eq:image} over frequency channels.

As derived in Appendix~\ref{app:passive_img}, the image variance decomposes as
\begin{equation}
  \Var\bigl[\Ihat(l,m)\bigr] = \alpha_0+\alpha_1+\alpha_2,
  \label{eq:alphas}
\end{equation}
where $\alpha_0$ represents intrinsic signal--signal fluctuations independent of the readout noise, $\alpha_1$ is the signal--noise cross term scaling linearly with $\sigma_i^2$ and $\sigma_j^2$, and $\alpha_2$ is the noise--noise term scaling as the product $\sigma_i^2\sigma_j^2$. These terms come from the products in $E_iE_j^*=S_iS_j^*+S_i\delta_j^*+\delta_iS_j^*+\delta_i\delta_j^*$: $\alpha_0$ from $S_iS_j^*$, which fluctuates because the passive signal is itself thermal noise, $\alpha_1$ from the products of signal and readout noise, and $\alpha_2$ from $\delta_i\delta_j^*$.

Under phase-squeezed readout, $\sigma_k^2\to g\sigma_k^2$ for all $k$, so $\alpha_1\to g\alpha_1$ and $\alpha_2\to g^2\alpha_2$, yielding
\begin{equation}
  \Var_{\rm sq}\bigl[\Ihat(l,m)\bigr] = \alpha_0+g\,\alpha_1+g^2\alpha_2.
  \label{eq:alphas_sq}
\end{equation}

Here we quantify image quality using the PSNR,
\begin{equation}
  \PSNR = 10\log_{10}\frac{I_{\max}^2}{\mathrm{MSE}},
  \label{eq:psnr}
\end{equation}
where $I_{\max}$ is the peak intensity of the reference image and
\begin{equation}
  \mathrm{MSE} = \E\Bigl[\bigl(\Ihat(l,m)-I_{\rm ref}(l,m)\bigr)^2\Bigr]
  \label{eq:mse}
\end{equation}
is the mean-squared error (MSE) of the reconstructed image $\Ihat(l,m)$ relative to a reference $I_{\rm ref}(l,m)$, averaged over the pixels of the evaluated region, with the expectation taken over repeated measurement realizations. The reference is the noise-free image $I_0(l,m)$, synthesized by Eq.~\eqref{eq:image} from the true visibilities $V_{ij}=\E[\Vhat_{ij}]$. Because $\Vhat_{ij}$ is unbiased (Appendix~\ref{app:passive_vis}), $\E[\Ihat(l,m)]=I_0(l,m)$ and the MSE equals $\Var[\Ihat(l,m)]$, so improvements in PSNR directly capture the reduction of image variance under phase-squeezed readout. For the measured images, $I_{\rm ref}$ is estimated from the other realizations of the same readout and $I_{\max}$ from the mean image of both readouts (Appendix~\ref{app:channels}). The gains of Figs.~\ref{fig:passive}(d) and~\ref{fig:active}(d) and those quoted in the text pool the MSE over all realizations of an operating point; the PSNRs of Figs.~\ref{fig:passive}(b) and~\ref{fig:active}(b) and the coherent and squeezed columns of Table~\ref{tab:active} are single-shot values.

The corresponding PSNR gain is
\begin{equation}
  \dPSNR = 10\log_{10}\frac{\alpha_0+\alpha_1+\alpha_2}{\alpha_0+g\,\alpha_1+g^2\alpha_2}.
  \label{eq:gain_passive}
\end{equation}
Because $\alpha_0,\alpha_1,\alpha_2\ge0$, the gain lies between 0 and $20\log_{10}(1/g)=2G$. The bound is twice the squeezing in decibels because the noise--noise term scales with the square of the readout-noise variance, as $g^2$; the readout noise itself falls only by $G$. It approaches $2G$ when the noise--noise term $\alpha_2$ dominates, vanishes when the scene term $\alpha_0$ dominates, and equals $G$ where $\alpha_0=g\,\alpha_2$. In a single frequency channel, for a spatially resolved sky observed by identical receivers, $\alpha_0:\alpha_1:\alpha_2=x^2:2x:1$, so that $\dPSNR=20\log_{10}[(x+1)/(x+g)]$, which decreases from $2G$ at $x\ll g$ through $G$ at $x=\sqrt g$ ($\xstarDb$~dB for $g=\gval$) to zero at $x\gg1$. On a point source, the source power adds coherently over the $M-1$ baselines of each receiver (self-noise~\cite{Kulkarni1989}), so point sources lose the advantage at a lower operating point than the diffuse background (Appendix~\ref{app:regimes}).

\subsection{Passive imaging experimental validation}\label{sec:passive_exp}

\begin{figure*}[t]
\includegraphics[width=\textwidth]{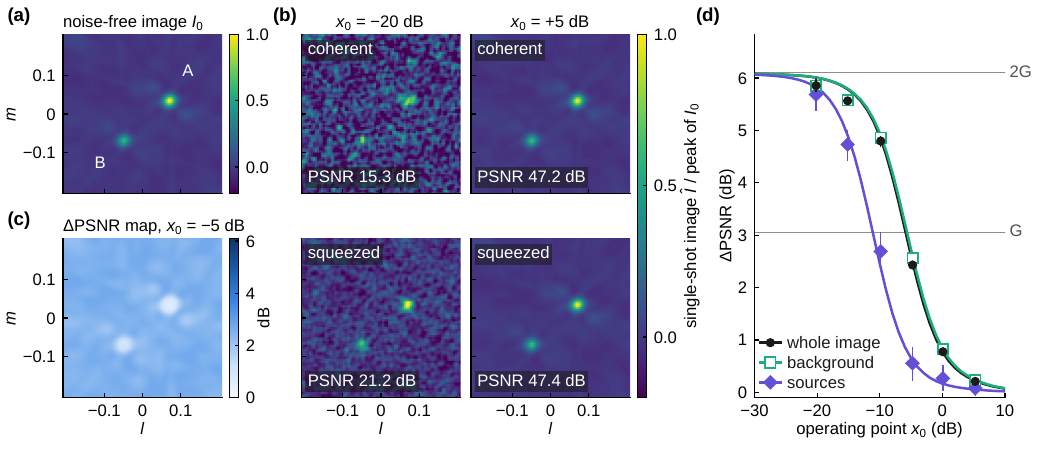}
\caption{Passive interferometric imaging with coherent and phase-squeezed readout ($M=24$ receivers, 5--20~MHz, $g=\gval$). (a)~Noise-free reference image $I_0(l,m)$, normalized to its peak, with sources A and B on faint diffuse emission. (b)~Single-shot reconstructions $\Ihat(l,m)$ with coherent (top) and phase-squeezed (bottom) readout at the operating points $x_0=\XlowDb$~dB (readout-noise dominated) and $\XhighDb$~dB (scene-noise dominated), on the color scale of (a); embedded values give the image PSNR. (c)~Pixel-wise PSNR gain predicted by Eq.~\eqref{eq:gain_passive} at $x_0=-5$~dB for a flat readout noise (in dB; the scale runs from 0 to $2G$). (d)~$\dPSNR$ versus $x_0$ for the whole image, the background ($|I_0|<5\%$ of the peak) and the sources ($I_0>50\%$ of the peak). Lines: theory, Eq.~\eqref{eq:gain_passive} with the variances summed over each region, at the measured receiver spectra, with equal signals and $G=\Gdb$~dB, with no free parameters. Symbols: measurements over \NskyPas\ independent sky realizations at the measured $x_0$, with 95\% intervals (Appendix~\ref{app:channels}). The same realizations are used at every $x_0$, so the points are correlated. Gray lines mark $G$ and $2G$. The axes $l$ and $m$ are direction cosines.}
\label{fig:passive}
\end{figure*}

In the experiment, passive RF imaging with multiple receivers is effectively implemented using a single balanced homodyne detector and time-division multiplexing (Appendix~\ref{app:channels}). Synthetic RF voltage waveforms---modeling the HF signals received by $M=24$ antennas distributed over a 1.6-km aperture---are generated numerically in the 5--20~MHz band as jointly Gaussian, zero-mean processes whose cross-correlations follow from the scene, including the geometric delays across the array. The scene comprises two point-like sources of unequal brightness (A and B) on a faint, clumpy diffuse ridge (barely visible on the linear color scale of Fig.~\ref{fig:passive}(a)), together with galactic emission that is resolved out by the array and modeled as uncorrelated between antennas; it serves as the ground truth for the noise-free reference image $I_0(l,m)$ [Fig.~\ref{fig:passive}(a)]. Each antenna waveform is sequentially replayed by the AWG, which drives the electro-optic phase modulator through a calibrated attenuator, thereby encoding each RF channel onto the phase quadrature of the optical carrier. The homodyne photocurrent is digitized for each waveform in turn on a clock shared with the AWG; sequential acquisition reproduces the readout-noise statistics of simultaneous multi-element reception (Appendix~\ref{app:channels}). Each repetition uses an independent sky realization, and the same realizations are replayed for both readouts. The RF level at the modulator, set by the attenuator, sets the operating point, which we vary in 5-dB steps from $x_0=-20$ to $+5$~dB. Appendix~\ref{app:scenes} describes how the sources and the background are emulated and lists the complete RF waveform parameters (Table~\ref{tab:scene_passive}).

Figure~\ref{fig:passive}(b) shows single-shot reconstructions $\Ihat(l,m)$ of the sky realization with the median $\dPSNR$, with coherent and phase-squeezed readout at two operating points. \ph{At $x_0=\XlowDb$~dB, where the image error is dominated by readout noise, the coherent image is covered by background fluctuations in which source B is barely visible; phase squeezing suppresses these fluctuations and raises the single-shot PSNR from \PsnrCohLow\ to \PsnrSqLow~dB. At $x_0=\XhighDb$~dB, the error is dominated by the intrinsic fluctuation $\alpha_0$ of the thermal scene, which squeezing cannot reduce: the two images are indistinguishable, with PSNRs of \PsnrCohHigh\ and \PsnrSqHigh~dB.} Figure~\ref{fig:passive}(c) maps the pixel-wise gain that Eq.~\eqref{eq:gain_passive} predicts at $x_0=-5$~dB for a flat readout noise: about \ThBgMfive~dB on the background and lower on the sources, where self-noise adds to $\alpha_0$. Figure~\ref{fig:passive}(d) summarizes the measured $\dPSNR$ versus $x_0$ together with Eq.~\eqref{eq:gain_passive} evaluated at the measured receiver spectra (lines). \ph{The whole-image gain decreases from \PfieldMtwenty~dB at $x_0=-20$~dB to \PfieldZero~dB at 0~dB and \PfieldPfive~dB at $+5$~dB.} \ifPasMeasCurves The predicted gain crosses $G$ at $x_0=\ph{\XcBgMeas}$~dB for the background, far below the single-channel value $\sqrt g$ ($\xstarDb$~dB), because $x(f)$ falls steeply across the band [Fig.~\ref{fig:spectra}(b)] and the low-frequency channels, where $x$ is largest, dominate the scene term $\alpha_0$ of the broadband image (Appendix~\ref{app:regimes}); for the sources the crossing moves \ph{\XcOffMeas}~dB lower, to \ph{$\XcSrcMeas$}~dB, because self-noise adds to $\alpha_0$ there.\else Equation~\eqref{eq:gain_passive} places the crossing of $G$ at $x_0=\XcBg$~dB for the background, below the single-channel value $\sqrt g$ because the sky brightness falls as $f^{-2.55}$ with RF frequency $f$~\cite{Cane1979} and the low-frequency channels, where $x$ is largest, dominate the scene term $\alpha_0$ of the broadband image (Appendix~\ref{app:regimes}); for the sources, the crossing moves to $\XcSrc$~dB, because self-noise adds to $\alpha_0$ there.\fi

\ph{Compared with the prediction for their own recordings, the whole-image and background gains agree to within \PasResHiMax~dB from $x_0=-10$~dB up. At $-20$ and $-15$~dB they lie $\PasResLoA\pm\PasCiLoA$ and $\PasResLoB\pm\PasCiLoB$~dB below it, because there the image error comes mostly from the top of the band, where the readout noise referred to the input is largest and the squeezing weakest (Fig.~\ref{fig:spectra}).\ifPasHaveGf\ With the measured frequency dependence of the squeezing, the predictions agree with every whole-image and background measurement to within \PasResGfMax~dB.\fi\ The gains of the sources agree with the prediction to within \PasSrcResMax~dB.}

\ph{Squeezing leaves the reconstructed scene unchanged: the brightness of the sources in the mean images of the two readouts agrees to within \SrcRatioHiMax\% from $x_0=-15$~dB up\ifMeanSigPas, and the small differences between the two mean images add at most \MeanMseFrac\% to the single-shot MSE\fi.}

\subsection{PSNR gain in active synthetic-aperture imaging}\label{sec:active}

In active synthetic-aperture imaging, a transmitter illuminates the scene with coherent RF waveforms---typically linear frequency-modulated (LFM) chirps---and echoes backscattered from the scene are coherently recorded as the platform traverses a synthetic aperture~\cite{Moreira2013,Cumming2005,Munson1989}. Here the channel index $k\in\{1,\dots,P\}$ labels individual aperture positions, with $P$ the total number of pulses; a frame is one complete synthetic-aperture acquisition of all $P$ echoes. Range resolution is achieved by matched filtering (range compression), which cross-correlates the record $E_k[n]$ of each pulse [Eq.~\eqref{eq:Ek}] with a replica $c(t)$ of the transmitted waveform and removes the RF carrier at its center frequency $f_c$,
\begin{equation}
  r_k(\tau) = e^{-i2\pi f_c\tau}\sum_n E_k[n]\,c^*(t_n-\tau),
  \label{eq:rc}
\end{equation}
where $t_n$ is the time of sample $n$, measured from the transmission of pulse $k$; an echo with the round-trip delay $\tau_0$ gives a peak of $|r_k(\tau)|$ at $\tau=\tau_0$, with the phase $-2\pi f_c\tau_0$. Cross-range (azimuth) resolution is achieved through coherent summation across aperture positions: for reconstruction pixel $q$, a point of the image grid on the ground (Table~\ref{tab:scene_active}), and aperture position $k$, the range-compressed return $r_k$ is evaluated at the round-trip delay $\tau_{k,q}$ of the pixel, phase-compensated by $e^{i\varphi_k(q)}$, and accumulated as
\begin{equation}
  z_q = \sum_k w_k(q)\,e^{i\varphi_k(q)}\,r_k(\tau_{k,q}) {}= z_{0,q}+n_q,
  \label{eq:bp}
\end{equation}
where $\varphi_k(q)=2\pi f_c\tau_{k,q}$ is the phase of the round trip at $f_c$ and $w_k(q)$ are apodization weights. The first term, $z_{0,q}$, comes from the echoes $S_k[n]$ of Eq.~\eqref{eq:Ek}: the deterministic noise-free signal determined by the scene reflectivity and imaging geometry. The second, $n_q$, is the additive noise term propagated from the raw readout noise $\delta_k[n]$ (Appendix~\ref{app:active_bp}). This time-domain back-projection operation coherently aligns contributions from all aperture positions, yielding a two-dimensional focused image of scene reflectivity.

Because all processing steps are linear in $E_k[n]$, the variance of $n_q$ follows directly from the readout-noise variance, here common to all pulses, $\sigma_k^2=\sigma^2$, and the focused-field noise variance is
\begin{equation}
  \Var(z_q) = \beta(q)\,\sigma^2,
  \label{eq:varz}
\end{equation}
where $\beta(q)$ is determined by pulse energy, aperture geometry, and reconstruction weights (Appendix~\ref{app:active_bp}). Under phase-squeezed readout, $\sigma^2\to g\sigma^2$, so $\Var_{\rm sq}(z_q)=g\,\beta(q)\,\sigma^2$.

The observable intensity image is $I_q=|z_q|^2=|z_{0,q}|^2+2\,\mathrm{Re}(z_{0,q}^*n_q)+|n_q|^2$. As derived in Appendix~\ref{app:active_img}, the intensity variance can be decomposed as
\begin{equation}
  \Var(I_q) = \gamma_1+\gamma_2,
  \label{eq:gammas}
\end{equation}
where $\gamma_1=2|z_{0,q}|^2\beta(q)\sigma^2$, the variance of the cross term $2\,\mathrm{Re}(z_{0,q}^*n_q)$, represents signal--noise cross terms proportional to $\sigma^2$, and $\gamma_2=\beta^2(q)\sigma^4$, the variance of the noise power $|n_q|^2$, represents noise--noise terms proportional to $\sigma^4$. Because the echo is deterministic, there is no counterpart of the scene term $\alpha_0$ of passive imaging; classical noise in the echoes and a fluctuation of the transduction add terms that squeezing does not reduce (end of this section). Under phase-squeezed readout, this becomes
\begin{equation}
  \Var_{\rm sq}(I_q) = g\,\gamma_1+g^2\gamma_2.
  \label{eq:gammas_sq}
\end{equation}
The PSNR compares $I_q$ with the noise-free intensity $|z_{0,q}|^2$, and the noise power $|n_q|^2$ also raises the mean intensity, $\E[I_q]=|z_{0,q}|^2+\beta(q)\sigma^2$ (Appendix~\ref{app:active_img}). The MSE is the variance plus this bias squared, and the squared bias, $\beta^2(q)\sigma^4$, equals $\gamma_2$:
\begin{align}
  \mathrm{MSE}(I_q) &= \Var(I_q)+\bigl(\E[I_q]-|z_{0,q}|^2\bigr)^2\nonumber\\
  &= (\gamma_1+\gamma_2)+\gamma_2 = \gamma_1+2\gamma_2,\nonumber\\
  \mathrm{MSE}_{\rm sq}(I_q) &= g\,\gamma_1+2g^2\gamma_2.
  \label{eq:mse_active}
\end{align}
The noise--noise term thus enters twice: as the fluctuation of the noise power about its mean (the $\gamma_2$ of the variance) and as that mean itself (the bias).
The corresponding PSNR gain is therefore
\begin{equation}
  \dPSNR = 10\log_{10}\frac{\gamma_1+2\gamma_2}{g\,\gamma_1+2g^2\gamma_2}.
  \label{eq:gain_active}
\end{equation}
In the noise-limited regime where $\gamma_2\gg\gamma_1$, the gain approaches $20\log_{10}(1/g)=2G$, again because $\gamma_2$ scales as the square of the readout-noise variance. In the signal-dominated regime where $\gamma_1\gg\gamma_2$, the gain approaches $10\log_{10}(1/g)=G$, so a deterministic target retains at least $G$ at any SNR when the classical noise of the echoes is negligible; classical noise in the echoes (sky or front-end noise) of power $x\sigma^2$, with $x$ defined as for a passive receiver (Sec.~\ref{sec:model}), replaces $g$ by $(x+g)/(x+1)$ (Appendix~\ref{app:active_ext}). A fluctuation of the RF-to-optical transduction from frame to frame, of relative amplitude variance $\kappa$, multiplies the echoes and adds, to leading order, $4\kappa|z_{0,q}|^4$ to the MSE of either readout; squeezing does not reduce it, so it lowers the gain where $|z_{0,q}|^2$ is large (Appendix~\ref{app:active_sq}).

\subsection{Active imaging experimental validation}\label{sec:active_exp}

In the experiment, active synthetic-aperture imaging is effectively implemented using the same single balanced homodyne receiver via time-division multiplexing (Appendix~\ref{app:channels}). LFM chirp echoes spanning 5--20~MHz are synthesized digitally from a deterministic scene model: for each of $P=250$ synthetic aperture positions, the round-trip delay and phase shift are precomputed from the platform trajectory and target geometry and encoded into the AWG waveform sequence. Each synthesized echo is sequentially replayed through the electro-optic phase modulator and optical readout stage. The RF level sets the echo SNR [Fig.~\ref{fig:spectra}(c)], the ratio of the echo power to the coherent readout-noise power within the 5--20~MHz band, both averaged over the receive window of each pulse; we vary it in 5-dB steps from $-35$ to $-10$~dB. Complete RF waveform parameters are listed in Table~\ref{tab:scene_active} (Appendix~\ref{app:scenes}).

\begin{figure*}[t]
\includegraphics[width=\textwidth]{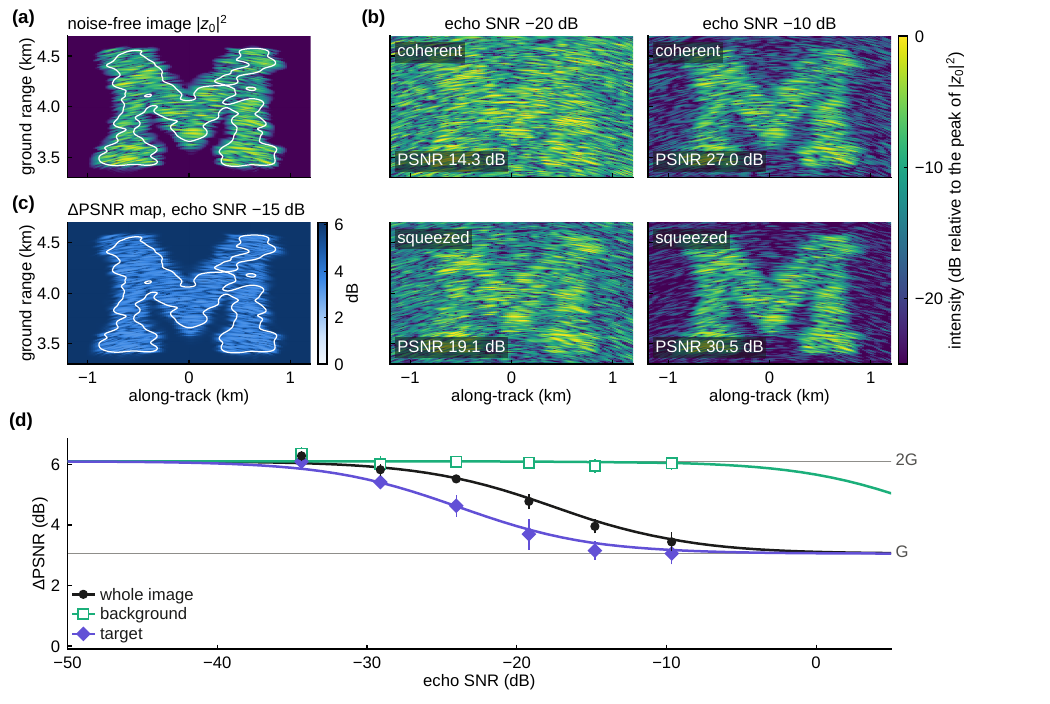}
\caption{Active synthetic-aperture imaging with coherent and phase-squeezed readout (5--20~MHz LFM chirps, $P=250$ aperture positions, $g=\gval$). (a)~Noise-free intensity image $|z_{0,q}|^2$ of the M-shaped target, in dB relative to its peak; the white contour outlines the target region of interest (ROI: $|z_{0,q}|^2>10\%$ of the peak; smoothed for display). (b)~Single-shot reconstructions with coherent (top) and phase-squeezed (bottom) readout at echo SNRs of $\EsnrLowDb$ and $\EsnrHighDb$~dB, in dB relative to the peak of $|z_{0,q}|^2$ at the same SNR, on the color scale of (a); embedded values give the image PSNR. (c)~Pixel-wise PSNR gain predicted by Eq.~\eqref{eq:gain_active} at an echo SNR of $-15$~dB, the SNR of Table~\ref{tab:active} (in dB; the scale runs from 0 to $2G$; white contour, the target ROI as in (a)). (d)~$\dPSNR$ versus echo SNR for the whole image, the background ROI (pixels more than 30~dB below the peak) and the target ROI. Lines: theory, Eq.~\eqref{eq:gain_active} summed over the pixels of each region, at the measured noise-free images and coherent image noise, with $G=\Gdb$~dB, with no free parameters. Symbols: measurements over \NframeAct\ frames at the measured echo SNR, with 95\% intervals (Appendix~\ref{app:channels}). Gray lines mark $G$ and $2G$.}
\label{fig:active}
\end{figure*}

\begin{table}[b]
\caption{PSNR for active imaging with coherent and phase-squeezed readout at an echo SNR of $-15$~dB (in dB). Coherent and squeezed: single-shot PSNR, mean $\pm$ standard deviation over the \NframeAct\ frames. $\dPSNR$: gain of the MSE pooled over the frames $\pm$ its 95\% confidence interval ($t$ interval from the gains of \ActNblocks\ batches of \ActBlock\ consecutive frames). Prediction: Eq.~\eqref{eq:gain_active} at the recording's measured inputs.}
\label{tab:active}
\begin{ruledtabular}
\begin{tabular}{lcccc}
Region & Coherent & Squeezed & $\dPSNR$ & Prediction\\
\hline
Whole image & $\ph{\TOvC}$ & $\ph{\TOvS}$ & $\ph{\TOvG\pm\TOvGpm}$ & \ph{\TOvP}\\
Target      & $\ph{\TTgC}$ & $\ph{\TTgS}$ & $\ph{\TTgG\pm\TTgGpm}$ & \ph{\TTgP}\\
Background  & $\ph{\TBgC}$ & $\ph{\TBgS}$ & $\ph{\TBgG\pm\TBgGpm}$ & \ph{\TBgP}\\
\end{tabular}
\end{ruledtabular}
\end{table}

Figure~\ref{fig:active}(b) shows single-shot reconstructed intensity images of the frame with the median $\dPSNR$, with coherent (top) and phase-squeezed (bottom) readout at echo SNRs of $\EsnrLowDb$ and $\EsnrHighDb$~dB, under identical RF illumination, platform trajectory, and reconstruction parameters. \ph{At $\EsnrLowDb$~dB, where readout noise dominates the whole image, the M-shaped target is barely discernible with coherent readout and emerges with phase-squeezed readout, which raises the single-shot PSNR from \PsnrActCohLow\ to \PsnrActSqLow~dB. At $\EsnrHighDb$~dB, the target is clear in both images, and squeezing mainly suppresses the noise of the background (PSNR \PsnrActCohHigh\ and \PsnrActSqHigh~dB). In both cases the target appears at the same location with the same spatial extent as in the noise-free image of Fig.~\ref{fig:active}(a), and averaged over the frames the complex images $z_q$ of the two readouts \ifMeanSigAct differ by slightly more than their noise\else agree within their noise, so squeezing leaves the point-spread function unchanged\fi.} Figure~\ref{fig:active}(c) maps the pixel-wise gain predicted by Eq.~\eqref{eq:gain_active}. In the noise-limited background far from the target, where the noise--noise term $\gamma_2\propto\sigma^4$ governs, squeezing produces the largest suppression, consistent with $g^2$ scaling. On the target, where the signal--noise cross term $\gamma_1\propto\sigma^2$ is significant, the improvement is more modest and approaches the $g$-scaling regime. This spatially varying gain, predicted by Eq.~\eqref{eq:gain_active} and tested region by region in Fig.~\ref{fig:active}(d) and Table~\ref{tab:active}, is the signature of the two-term noise structure derived in the model.

To quantify these observations, we define two regions of interest (ROIs) in the image plane. The target ROI contains the pixels where the noise-free intensity $|z_{0,q}|^2$ exceeds 10\% of its peak [outlined in Fig.~\ref{fig:active}(a)]. The background ROI contains the weak-signal pixels, more than 30~dB below the peak. Both ROIs are defined on the noise-free reference image and applied identically to all realizations and both readout conditions.

Figure~\ref{fig:active}(d) follows the gains as the echo SNR increases, with Eq.~\eqref{eq:gain_active} at the measured noise-free images and coherent image noise as lines. \ph{The background gain stays near $2G$ at every echo SNR (\ActBgMin--\ActBgMax~dB; above $2G$ at $-35$~dB, where the squeezing of that recording, referred to its signal, was \GeffLow~dB), while the target gain falls from \ActTgtFirst~dB at $-35$~dB to \ActTgtLast~dB at $-10$~dB, close to $G$, as the lines predict. Compared with the prediction for each recording's own squeezing level, the background and the whole image agree to within \ActBgResMax\ and \ActFieldResMax~dB at every echo SNR and the target to within \ActTgtResMax~dB\ifActAllIn, each inside the 95\% interval of its measurement\fi. The measured fluctuation of the squeezed carrier's transduction, which squeezing does not reduce, would lower the target gain by up to \ActWanderEffMax~dB.} Table~\ref{tab:active} reports the PSNR over \NframeAct\ frames for three evaluation regions, each probing a different regime of the measurement model. The whole-image PSNR is computed over the full reconstruction grid against the noise-free reference image; \ph{squeezing improves it by \TOvG~dB}. The target ROI covers the M-shaped target, where the local SNR is moderate (median $|z_{0,q}|^2/\Var(z_q)\approx\RhoTgt$) and the $\gamma_1$ term dominates; \ph{its gain of \TTgG~dB is close to this recording's lower bound of the signal-dominated regime, $G_{\rm eff}=\GeffTable$~dB (predicted: \TTgP~dB)}. The background ROI isolates the noise floor, where $\gamma_2$ dominates; \ph{its gain of \TBgG~dB matches the \TBgP~dB predicted for this recording, at its upper bound of the noise-limited regime, $2G_{\rm eff}=\TwoGeffTable$~dB}.

Across both experiments, \ph{the model reproduces the measured PSNR gains without free parameters, to within \FBResAllMax~dB for the whole images and backgrounds, \ActTgtResMax~dB on the active target and \PasSrcResMax~dB on the passive point sources. With the measured frequency dependence of the squeezing, \ifPasGfAllIn every passive prediction lies inside the 95\% interval of its measurement\else the passive predictions agree with the measurements to within \PasResGfMax~dB\fi; the deviations of the active gains trend slightly with the echo SNR.} The enhancement is governed by a single receiver-level parameter, $g$, which rescales the additive readout noise in both modalities, while the resulting image gain depends on the estimator and on the operating point.

\section{Discussion}\label{sec:discussion}

This work establishes a receiver-level strategy for quantum enhancement in RF-photonic sensing. Encoding RF information onto a bright phase-squeezed optical carrier and reading it out by phase-sensitive homodyne detection reduces the additive measurement noise at the optical readout stage by a factor $g<1$~\cite{Caves1981,Yuen1983,Haus2004}. Within the unified measurement model developed here, squeezing acts as a multiplicative rescaling of the variance of the additive noise term in the complex samples. Because active synthetic-aperture focusing is linear and passive interferometric reconstruction bilinear in the measured fields, this variance reduction propagates predictably through both imaging pipelines. The resulting noise reduction is governed by a single experimentally measurable parameter $g$, independent of sensing modality; how much of it appears in the image is set by the share of readout noise in the image error.

Conceptually, this architecture differs fundamentally from entanglement-based radar and quantum illumination protocols~\cite{Tan2008,Barzanjeh2015,Assouly2023,Zhang2015}, which rely on nonclassical correlations between transmitted and retained modes, require operating at low probe power to maintain quantum advantage, and often demand quantum storage or joint measurements of the returned and retained modes~\cite{Shapiro2020}. In contrast, the RF field here remains classical, and all quantum resources are localized at the optical readout stage. This receiver-centric deployment of squeezing parallels strategies used in precision optical metrology~\cite{Giovannetti2011,Lawrie2019,Xiao1987,Atkinson2021}, where quantum states of light are engineered to suppress measurement noise without modifying the signal source itself. Consequently, the quantum resource is not exposed to RF propagation loss or environmental decoherence, provided optical loss within the receiver is controlled; classical RF loss and noise only set the operating point. \ph{The measured gains follow the predicted dependence on the operating point without free parameters, and apart from the larger fluctuation of the squeezed carrier's transduction (below) and the interference lines of the coherent records (Appendix~\ref{app:channels}), the two readouts differ only in their readout noise.}

The benefit of squeezing is greatest when optical readout noise constitutes a significant fraction of the total system noise budget, consistent with general quantum metrology principles~\cite{Giovannetti2011,Lawrie2019}. In signal-dominated regimes or in the presence of strong classical RF fluctuations, the relative improvement diminishes, in agreement with the variance decomposition derived in Appendices~\ref{app:passive} and~\ref{app:active}. In passive HF imaging, this is the case for RF-photonic receivers without electronic gain ahead of the modulator, whose readout noise referred to the input can exceed even the bright galactic background (Appendix~\ref{app:x0}). The regime is not specific to HF: the sky brightness falls steeply with frequency, from about $10^5$~K at 12.5~MHz to a few kelvin at 1--10~GHz~\cite{ITU372}, while $T_{\rm q}$, set by the modulator and the carrier power, does not fall with it, so at higher frequencies such receivers operate at $x\ll g$, where the passive gain approaches $2G$ and the active gain lies between $G$ and $2G$. Like an RF amplifier, the RF-photonic receiver adds noise of its own, of temperature $T_{\rm q}/G_{\rm fe}$ referred to the antenna. Squeezing lowers it by the factor $g$, as a carrier power raised to $P_{\rm opt}/g$ would (Sec.~\ref{sec:model}), whereas an amplifier ahead of the modulator lowers it by raising $G_{\rm fe}$ but adds noise of its own, $T_N$, and lowers the largest input handled linearly by the same factor. Equivalently, squeezing shortens the integration time needed for a given image quality. In passive imaging, where the image error falls inversely with the integration time for independent samples, it does so by the factor $10^{\dPSNR/10}$ (at most $1/g^2$). In active imaging, coherent averaging of $K$ complex images of a fixed, phase-stable scene, before the intensity is formed, makes every image statistic depend on $\sigma^2/K$ alone, so squeezing shortens the acquisition by the factor $1/g$ (Appendices~\ref{app:regimes} and~\ref{app:active_ext}). This matters most where the integration time cannot be extended, as for time-variable scenes or a synthetic aperture of fixed length. Because the quantum resource is confined to the optical receiver, the architecture is naturally extendable to broadband, multi-static, and distributed RF-photonic sensing systems such as large-scale interferometric arrays~\cite{Hotan2021}.

The readout noise temperature itself scales as $V_\pi^2/P_{\rm opt}$, which sets the classical path to lowering it~\cite{Cox2006}. The free-space modulator of the experiment gives $T_{\rm q}\approx\TqMeas$~K at 20~$\upmu$W (Appendix~\ref{app:readout}); integrated thin-film lithium-niobate modulators reach half-wave voltages of a few volts or less~\cite{Wang2018}, and for the integrated phase modulator assumed in Appendix~\ref{app:x0} ($V_\pi=3.5$~V), $T_{\rm q}$ falls to $1.3\times10^5$~K at 1~mW and $1.3\times10^4$~K at 10~mW. The carrier power is in turn limited by the power handling of the modulator and the detectors and, in a distributed array, by the optical power budget; at a given carrier power, squeezing lowers $T_{\rm q}$ by the further factor $g$, which improves the image most where $T_{\rm q}$ exceeds the noise temperature of the scene ($x<1$). Squeezing is therefore most valuable for unpowered photonic front ends, a modulator at the antenna connected only by optical fiber, where a low-noise amplifier right after the antenna is practically unavailable. Example use cases include RF pickups in radiation areas with restricted maintenance access, such as accelerators~\cite{Arteche2019} and nuclear and fusion facilities~\cite{Girard2013,Nishiura2017}; permanent receivers in hot, inaccessible boreholes, such as geothermal wells and thermal-recovery reservoirs~\cite{Sato2007,Reinsch2013,Goiffon2006}; electrically floating sensors inside high-voltage equipment, such as cable joints and switchgear, for insulation diagnostics~\cite{Pommerenke2000,Ryu2007}; RF and microwave sensors at a heat-limited cryogenic stage~\cite{Youssefi2021,Arnold2025}; miniature MRI receive probes and catheter-mounted coils, whose conducting cables would heat tissue~\cite{Fandrey2008,Ayde2014}; and unpowered antenna nodes of fiber-distributed arrays~\cite{Sato2007,Wake1997}.

Three features of the present implementation are important for how the results carry over to a deployed receiver. First, the antenna array and the synthetic aperture are emulated. A single receiver records the channels one after another, which reproduces the readout-noise statistics of simultaneous reception because the readout noise of different time slots is independent (Appendix~\ref{app:channels}). This leaves aside the channel-to-channel calibration that a physical array requires, and interference picked up by the single receiver repeats in every time slot and so correlates the emulated channels, as the interference lines in the records did (Appendix~\ref{app:channels}). Second, the RF-to-optical transduction of the squeezed carrier fluctuated on time scales of tens of seconds, by up to \ph{\WanderActSqPowMax\%} in power within an active recording, against at most \ph{\WanderActCohPowMax\%} for the coherent carrier; we removed it from the passive images using the known replayed waveforms (Appendix~\ref{app:channels}), and in active imaging it would lower the target gain by up to \ph{\ActWanderEffMax}~dB. A deployed receiver has no replayed reference and would instead stabilize the carrier power or track the transduction with a pilot tone outside the imaging band. Third, the squeezing weakens toward the top of the band [Fig.~\ref{fig:spectra}(a)], which costs up to \ph{\PasResLoB}~dB of passive gain at the lowest operating points (Sec.~\ref{sec:passive_exp}); a flatter squeezing spectrum would recover it.

Looking forward, further improvement is achievable by increasing the squeezing at the output of the optical parametric amplifier (OPA). State-of-the-art squeezed-light sources have demonstrated noise reductions of up to 15~dB~\cite{Vahlbruch2016}, corresponding to $g\approx0.03$, which would yield PSNR gains approaching $2G=30$~dB where the readout noise dominates, provided the readout noise detected in the receiver is reduced as much; this requires nearly lossless optics between source and detector and negligible electronic noise, both of which enter $g$ (Sec.~\ref{sec:model}). That regime shrinks as $g$ falls, because the gain approaches $2G$ only for $x\ll g$ and has already fallen to $G$ at $x=\sqrt g$, which is $-7.5$~dB for 15~dB of squeezing. In the present implementation, optical loss between the OPA and the homodyne detector is the primary factor limiting $g$ to \gval, below the squeezing levels reported for RF-signal estimation~\cite{Li2024}; more squeezing enters the model only through $g$. Optical loss on the way to the detector, including fiber distribution to remote antenna nodes, degrades $g$ in the same way. Advances in low-loss integrated photonics~\cite{Ji2024,Liu2025,Wang2024}, on-chip squeezed-light phase sensing~\cite{Stokowski2023}, and integrated high-bandwidth homodyne detection~\cite{Bruynsteen2021} offer a direct path to closing this gap. A squeezed carrier split among $M$ receivers leaves each receiver, read out alone, with $1-(1-g)/M$, whereas the combination of their outputs matched to the splitting, such as a joint function of the output RF waves across the distributed sensor network, chosen by the optical network's splitting ratios and phase shifts, keeps the full $g$ of the source, as in distributed quantum sensing~\cite{Zhuang2018,Guo2020,Xia2020}; keeping the full $g$ at every pixel of an image at once takes as many squeezed modes as receivers. The squeezed readout is also not tied to the band studied here: a cavity OPA squeezes the sidebands around every multiple of its free spectral range~\cite{Senior2007,Chen2014}, and single-pass waveguide OPAs squeeze over terahertz bandwidths~\cite{Kashiwazaki2020,Shaked2018,Nehra2022}, so the same readout extends to gigahertz RF bands with a homodyne detector of matching bandwidth~\cite{Bruynsteen2021}. Squeezing has already been read out in real time from DC to 43~GHz~\cite{Inoue2023}. Beyond this, a physically distributed multi-antenna array in lieu of the time-division multiplexing used in the present experiment constitutes a pathway toward practical quantum-enhanced RF-photonic imaging systems. Moreover, the combination of receiver-level quantum enhancement and the low-loss optical fiber networking capabilities of RF-photonic platforms~\cite{Hotan2021,Dong2020} positions this approach as a promising route toward quantum-enhanced large-aperture passive and active RF imaging systems.

\begin{acknowledgments}
This work was supported by the Department of Energy ARPA-E (DE-AR0002067), Halliburton Technology Partners LLC, the Office of Naval Research (N00014-23-1-2296), and the National Science Foundation (2317471 and 2326746). V.M. acknowledges a scholarship for doctoral studies from the Faculty of Engineering and the Department of Electrical Engineering, Kasetsart University (Thailand).

Claude (Anthropic, Opus 5.5) assisted with language editing and drafting portions of the manuscript. The authors directed these uses, reviewed and revised all AI-assisted text, verified its scientific accuracy, and take full responsibility for the content of this article.
\end{acknowledgments}

\section*{Data availability}
The data that support the findings of this article are not publicly available. The data are available from the authors upon reasonable request.

\appendix

\begin{figure}[t]
\includegraphics[width=\columnwidth]{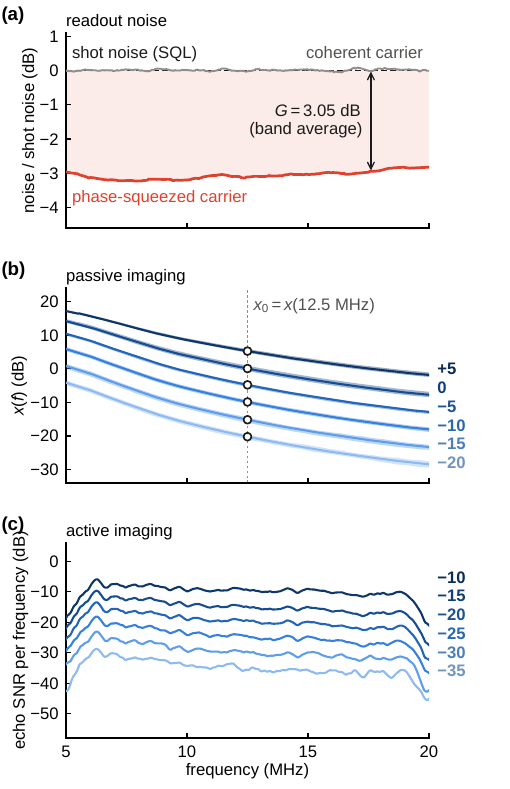}
\caption{Receiver characterization. (a)~\ph{Measured homodyne noise relative to the shot noise of the local oscillator: the coherent carrier (spectrum analyzer, smoothed over 0.5~MHz) and the phase-squeezed carrier at the same power as the digitized records show it}; $G$ is the band average (Sec.~\ref{sec:receiver}). (b)~Passive imaging: signal-to-readout-noise ratio $x(f)$ of the replayed sky at the six operating points, labeled by $x_0$ in dB in the colors of their curves (mean over the 16 recordings of each operating point; shading, their range); $x_0$ is the value at the 12.5-MHz band center (circles). (c)~Active imaging: echo SNR per frequency, the ratio of the echo's power spectral density to that of the coherent readout noise, at the six echo SNRs (labels, in dB), smoothed over 0.4~MHz; the peaks near the band edges are the Fresnel ripples of the chirp spectrum.}
\label{fig:spectra}
\end{figure}

\section{Experimental setup}\label{app:setup}

We realize the quantum-enhanced receiver architecture of Fig.~\ref{fig:architecture} using a sub-threshold OPA-based squeezed-light source, electro-optic phase encoding, and balanced homodyne detection. The same optical readout stage is used for both passive and active imaging experiments, with the two modalities differing only in the RF waveforms synthesized and replayed through the AWG.

\subsection{Optical source and squeezed-state generation}\label{app:source}

A continuous-wave laser at 1550~nm serves as the master optical source. The output is amplified by an erbium-doped fiber amplifier and spatially filtered through mode-cleaning cavities to suppress higher-order spatial modes and reduce technical noise. The filtered beam is split into two arms: a local-oscillator (LO) arm and a pump-generation arm. In the pump arm, the 1550-nm light is frequency-doubled via second-harmonic generation in a periodically poled potassium titanyl phosphate (PPKTP) crystal, producing 775-nm pump light. This pump drives a sub-threshold OPA containing a type-0 PPKTP crystal, together with a 1550-nm seed beam derived from the same master laser, generating a bright phase-squeezed field at 1550~nm. Its squeezing bandwidth encompasses the 5--20~MHz RF band; the squeezing level in the detected records is given in Sec.~\ref{sec:receiver} and Fig.~\ref{fig:spectra}(a).

\subsection{Electro-optic transduction and homodyne detection}\label{app:readout}

\begin{figure}[t]
\includegraphics[width=\columnwidth]{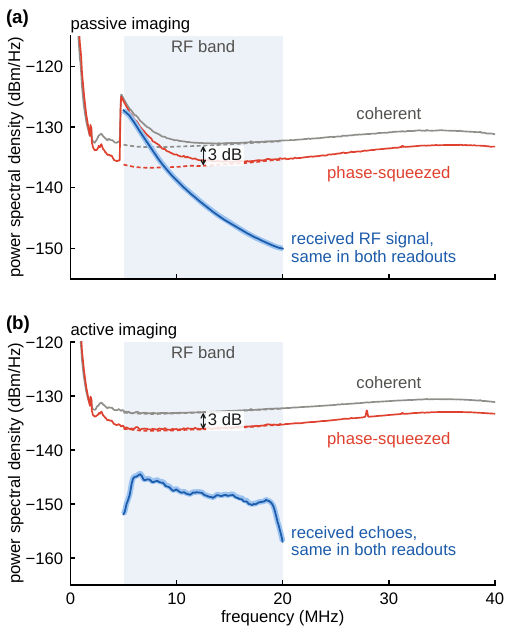}
\caption{Received recording pairs: the power spectral density into the digitizer's 50-$\Omega$ input before any processing, with coherent (gray) and phase-squeezed readout (red). (a)~Passive imaging, the recording of one data set at $x_0=\ph{\RecXzero}$~dB. (b)~Active imaging, the recording at an echo SNR of $\ph{\RecActSnr}$~dB.\ifRecMatched\ The coherent records are the references recorded at the squeezed records' signals (Appendix~\ref{app:channels}).\fi\ In the 5--20-MHz RF band (shaded), the projection onto the replayed waveforms splits each record into the received signal (blue; coherent, wide light line; squeezed, narrow dark line) and the readout noise (dashed); arrows mark the band-averaged reduction of the readout noise. \ph{Below 3~MHz, low-frequency technical noise, including that of the phase locks, dominates both records.} Only the band enters the processing.}
\label{fig:record}
\end{figure}

RF signals in the 5--20~MHz band are synthesized digitally and replayed directly, without frequency conversion, by the AWG (Tektronix AWG7102, 8~bits, 100~MS/s) at a fixed amplitude. A current-controlled attenuator (Mini-Circuits ZAS-3+) sets the RF level, and thereby the operating point $x_0$ (passive imaging) or the echo SNR (active imaging), in 5-dB steps. Its harmonic distortion puts the in-band intermodulation products of the broadband waveforms \ph{more than 25~dB below the readout noise}. The attenuator drives a free-space electro-optic phase modulator (Thorlabs EO-PM-NR-C3, terminated with a 50-$\Omega$ feed-through), which encodes the RF waveform onto the phase quadrature of the optical carrier. The carrier has a power $P_{\rm opt}$ of about \ph{20}~$\upmu$W for both readouts. Referred to the modulator input, the coherent readout noise measured in the records corresponds to $T_{\rm q}\approx\ph{\TqMeas}$~K, which for this carrier power and $\eta=0.9$ implies $V_\pi\approx\ph{\VpiImplied}$~V at 1550~nm; the replayed RF levels thus emulate an antenna behind a noiseless gain of \ph{\GfeLo--\GfeHi}~dB (Appendix~\ref{app:x0}). For squeezed readout, the bright phase-squeezed output of the OPA is directed to the signal port of the homodyne beamsplitter; for coherent readout, coherent light from the same source at the same carrier power replaces it, with all acquisition and processing parameters unchanged\ifmatched{} except the RF level of each coherent recording, which was matched to the signal of its squeezed counterpart (Appendix~\ref{app:channels})\fi. 

Optical readout is performed using balanced homodyne detection. The carrier interferes with the LO (40~mW) on a 50:50 beamsplitter, the two output ports are detected by a matched pair of photodiodes, and the relative phase is stabilized to the phase quadrature by a feedback servo acting on a piezoelectric actuator. The difference photocurrent is amplified and split between an oscilloscope (Teledyne LeCroy WavePro 804HD, 12~bits, 500~MS/s), whose sample clock is locked to the AWG, and an electrical spectrum analyzer (Agilent E4440A) that measures the noise levels. The shot noise exceeds the electronic noise of the digitizer by \ph{\ScopeClearDb}~dB over the band.

The operating point $x(f)$, the readout-noise variance $\sigma_k^2$ and the squeezing factor $g$ of Eq.~\eqref{eq:sq} are obtained from the homodyne records themselves. Each record is projected onto the known replayed waveform, with one transfer function $H(f)$ per recording estimated by least squares over all its time slots; the projection gives the received signal, and the residual is the readout noise. Figure~\ref{fig:record} shows this split for a recording pair of each experiment: the received signals of the two readouts differ by at most \ph{\RecSigMax}~dB. $x(f)$ is the ratio of the replayed signal's spectrum to the residual's spectrum, both referred to the modulator input through $|H(f)|^2$. $g$ is the ratio of the in-band residual powers of the squeezed and coherent records of the same data set and operating point, with the coherent records of the first coherent run (Appendix~\ref{app:channels}), averaged over all \ph{\NrecPairs} pairs: \ph{$G=\GrecDb$~dB}. \ifGfRecords For the predictions with a frequency-dependent $g$ (Sec.~\ref{sec:passive_exp}), we take its shape from the ratio of the residual power spectra of the two readouts in one data set at $x_0=-10$~dB, scaled so that its average, weighted by the noise spectrum of the records, equals $g$ [Fig.~\ref{fig:spectra}(a)].\else For the predictions with a frequency-dependent $g$ (Sec.~\ref{sec:passive_exp}), we take its shape from the analyzer, with its own noise floor removed and that of the digitizer added, smoothed over 1~MHz and scaled so that its average, weighted by the noise spectrum of the records, equals $g$; $G(f)$ then falls from \ph{\GfLoDb}~dB at 5--8~MHz to \ph{\GfHiDb}~dB at 18--20~MHz.\fi

\subsection{Synthesized RF scenes}\label{app:scenes}

Tables~\ref{tab:scene_passive} and~\ref{tab:scene_active} list the parameters of the passive and active scenes. For passive imaging, $M=24$ antenna waveforms of 1~ms duration are synthesized in software as jointly wide-sense-stationary, zero-mean, real Gaussian voltage waveforms occupying the 5--20~MHz band, with the cross-correlations and true geometric delays of the scene. The sky is modeled as a set of point emitters: the two sources, each at a single direction, and the diffuse ridge, a band with a Gaussian cross-section, modulated by fixed random clumps and divided into $30\times30$ patches. Each emitter radiates independent Gaussian noise with the power spectrum of its antenna temperature, and the voltage of each antenna is the sum of these fields with the true geometric delay from the emitter to that antenna. The galactic emission, smooth all-sky emission that the array resolves out, is added as noise independent at each antenna. At 12.5~MHz, source~A, source~B, the ridge, and the galactic emission contribute $1:0.6:1:5.2$ of the power of each antenna, so that the two sources carry about one fifth of the antenna temperature; the RF level then scales the whole waveform to the operating point $x_0$. The sources have no extent of their own: their width in the images is the point-spread function of the array. Every repetition uses an independent sky realization, a new draw of the noise of every emitter, replayed for both readouts; replaying the same realization would turn the scene term $\alpha_0$ into a fixed bias that averaging over repetitions does not sample.

For active imaging, LFM chirps sweeping 5--20~MHz in 20~$\upmu$s are synthesized for each of $P=250$ aperture positions; round-trip delays to every scatterer of the scene are precomputed from the platform trajectory and scene geometry (Table~\ref{tab:scene_active}) and encoded into the AWG waveform sequence. The pulse repetition frequency and the platform motion of Table~\ref{tab:scene_active} are those of the emulated radar; in the replay, the $P$ echoes of a frame follow one another within one record of 13.7~ms.

\begin{table*}[t]
\caption{Parameters for passive interferometric imaging.}\label{tab:scene_passive}
\begin{ruledtabular}
\begin{tabular}{@{}p{0.27\textwidth}p{0.68\textwidth}@{}}
Parameter & Value\\
\hline
RF band & 5--20~MHz\\
Band center (reference for $x_0$) & 12.5~MHz\\
Number of receivers $M$ & 24, random in a 1.6-km disk (minimum spacing 40~m)\\
Image & direction cosines $|l|,|m|\le\sin12^\circ$ (field $24^\circ$), $81\times81$ pixels; resolution $0.54^\circ$ at 20~MHz\\
Scene & point sources A and B at $(l,m)=(\sin4^\circ,\sin2^\circ)$ and $(-\sin2.9^\circ,-\sin4^\circ)$, power ratio 1:0.6, antenna temperature $\propto f^{-2.8}$; a diffuse ridge with the power of source A, of Gaussian cross-section with a standard deviation of $2.5^\circ$; galactic emission resolved out by the array, modeled as uncorrelated between receivers, with twice the correlated in-field power; diffuse brightness $\propto f^{-2.55}$\\
Sampling and channels & 100~MS/s; 4096-sample blocks (24.4~kHz channels, 615 in band)\\
Integration time per receiver $T_{\rm int}$ & 1~ms ($N=24$ samples per channel, $BT_{\rm int}=1.5\times10^{4}$)\\
Weights & natural baseline weighting ($w_b=1$), uniform channel weights\\
Squeezing factor $g$ & 0.501 (3~dB) in the Monte Carlo simulation (Appendix~\ref{app:mc}); the comparisons with the measurements use the measured $g$ of each recording\\
Operating point $x_0$ & $-20$ to $+5$~dB\\
Statistical realizations\newline per condition & 208 independent skies, the same skies for both readouts and at every $x_0$\\
\end{tabular}
\end{ruledtabular}
\end{table*}

\begin{table*}[t]
\caption{Parameters for active synthetic-aperture imaging.}\label{tab:scene_active}
\begin{ruledtabular}
\begin{tabular}{@{}p{0.27\textwidth}p{0.68\textwidth}@{}}
Parameter & Value\\
\hline
RF band & 5--20~MHz\\
Chirp bandwidth $B$ & 15~MHz (linear frequency modulation, 5 to 20~MHz), 10\% Tukey taper\\
Pulse duration $T_p$ & 20~$\upmu$s\\
Chirp rate $K_r=B/T_p$ & $7.5\times10^{11}$~Hz/s\\
Sampling & 100~MS/s; receive window 27.3--61.9~$\upmu$s after each pulse; range-compressed data upsampled $\times4$ before interpolation\\
Pulse repetition frequency & 30~Hz\\
Number of pulses $P$ & 250 (4~m spacing, aperture 1~km)\\
Platform velocity & 120~m/s\\
Platform altitude & 3~km\\
Propagation & two-way spherical spreading (echo amplitude $\propto$ range$^{-2}$)\\
Scene & M-shaped target, $1.7\times1.15$~km at 4~km ground range, represented by unit-amplitude scatterers with fixed random phases on a $30\times8$~m grid, identical in every frame (deterministic)\\
Image & $2.4\times1.4$~km, pixels of 20~m (along track) $\times$ 6~m (ground range); Hann apodization $w_k(q)$ across the aperture\\
Squeezing factor $g$ & 0.501 (3~dB) in the Monte Carlo simulation (Appendix~\ref{app:active_ext}); the comparisons with the measurements use the measured $g$ of each recording\\
Echo SNR & $-35$ to $-10$~dB: echo power over the coherent readout-noise power within 5--20~MHz, during the receive window\\
Statistical realizations\newline per condition & 100 frames in the measurement (one frame: all $P$ echoes of one synthetic aperture), the same frames for both readouts; 200 in the Monte Carlo simulation\\
\end{tabular}
\end{ruledtabular}
\end{table*}

\subsection{RF channel implementation and signal processing}\label{app:channels}

Both imaging modalities are implemented using time-division multiplexing of the single physical homodyne receiver, allowing the receiver architecture to be validated in isolation from multi-channel hardware: the RF waveforms of the different receivers (passive) or aperture positions (active) are sequentially replayed through the AWG and applied to the modulator one after another. For passive imaging, the $M$ waveforms of one sky realization are synthesized with the joint statistics of Eq.~\eqref{eq:c_Vij} (Appendix~\ref{app:scenes}) and replayed in successive time slots; they carry the joint signal statistics that simultaneous receivers would record. The readout noise of different slots comes from disjoint time intervals, longer than its correlation time ($\sim1/B$, with $B=15$~MHz the bandwidth), and is therefore independent, as for physically separate receivers [Eq.~\eqref{eq:c_noisevar}]; \ph{in the records, the residuals of different time slots are uncorrelated, apart from a few narrow interference lines, mainly in the coherent records, whose source was not identified: for one data set at $x_0=-10$~dB, the cross-spectra between the residuals of different slots, averaged over its 13 realizations, have \SlotRatioSq\ (squeezed) and \SlotRatioCoh\ (coherent) times the power expected for independent noise, and \SlotRatioSqClean\ and \SlotRatioCohClean\ ($\pm\SlotErr$) without the 10 strongest of the 615 channels. Treated as extra readout noise in their channels, the lines (\SlotLinesCoh~MHz) change the predicted gains by less than 0.02~dB}. Sequential replay thus reproduces every term of Eqs.~\eqref{eq:c_covV} and~\eqref{eq:alphas}, provided the replay and the digitizer share one clock (so that the relative delays between receivers are preserved) and the readout conditions ($P_{\rm opt}$, local-oscillator phase, $g$) stay constant over the slots of a realization. For active synthetic-aperture imaging with a deterministic scene, the echo delays and phases are precomputed from the radar geometry, so sequential replay reproduces the intended pulse-by-pulse measurement exactly, and the readout noise is independent across pulses. Sequential acquisition does not reproduce what a physical array adds: channel-dependent transfer functions and their calibration, the drifts of independent oscillators, and environmental noise correlated between receivers.

Each run, squeezed or coherent, records the 16 passive data sets of 13 sky realizations at their six operating points and \NframeAct\ frames of the active scene at each of its six echo SNRs.

Both readouts must carry the same RF signal: because each readout's images are referred to its own measured signal, a squeezed-readout signal power smaller by a fraction $\zeta$ lowers the gain by up to $-20\log_{10}(1-\zeta)$~dB, 0.45~dB for $\zeta=5\%$. \ifmatched Because the power of the squeezed carrier drifted during the squeezed run, we recorded the coherent reference a second time, with the attenuator of every recording set to the signal measured in its squeezed counterpart, which brought the two signals to within \ph{\TafterRms\%} (rms); Figs.~\ref{fig:passive} and~\ref{fig:active} and Table~\ref{tab:active} use this reference, and the predictions use the squeezing level of each recording referred to its signal, $g_{\rm eff}=(\sigma_{\rm sq}^2/\sigma_{\rm coh}^2)/(P_{\rm sq}/P_{\rm coh})$, where the two ratios are those of the in-band readout-noise and signal powers of the squeezed and of the matched coherent recording. These predictions lie within \PasLineDevMax~dB (passive) and \ActLineDevMax~dB (active) of the lines of Figs.~\ref{fig:passive}(d) and~\ref{fig:active}(d), which use the average $G$. \fi The noise-free images $I_0$ and $z_{0,q}$ follow from the known scenes through the measured response; every other input of the predictions is measured in each recording's own records.

Digitized homodyne records are resampled to 100~MS/s with a digital anti-alias filter, band-passed to 5--20~MHz, aligned to the AWG timing, and processed directly at RF. For passive imaging, each record is divided into blocks and Fourier transformed before the cross-multiplication of receiver pairs (FX correlation); visibilities $\Vhat_{ij}$ are estimated channel by channel [Eq.~\eqref{eq:vis}], and the image is formed by the inverse Fourier sum of Eq.~\eqref{eq:image}, summed over channels (Table~\ref{tab:scene_passive}). Every channel is divided by the measured $|H(f)|^2$ of its recording, so that both readouts give images in units of the replayed RF signal with the uniform channel weights of the model, and every sky realization is divided by its own signal level, which removes the fluctuation of the transduction from one realization to the next. Active echoes are converted to analytic signals, range-compressed by matched filtering [Eq.~\eqref{eq:rc}], and focused via time-domain back-projection [Eq.~\eqref{eq:bp}]; the images of the squeezed readout are referred to its own signal through the amplitude ratio of the recording's signals.

The PSNR reference of the measured passive images is, for each realization, the mean image of the other realizations of the same readout at the same operating point (leave-one-out), which raises the MSE of both readouts by the same factor and leaves $\dPSNR$ unchanged; $I_{\max}$ is the peak of the mean image of both readouts. The active reference is $|z_{0,q}|^2$, obtained by passing the noise-free echoes through the measured system response and the same reconstruction. The 95\% intervals of the passive gains are percentile intervals of 1000 bootstrap resamples of the 16 recordings, with the prediction evaluated anew for every resample; the active intervals are $t$ intervals about the pooled gain, with the standard error of the gains of \ActNblocks\ batches of \ActBlock\ consecutive frames (about 20~s each), which allow for the slow fluctuation of the squeezed carrier's transduction. With one recording per echo SNR, the intervals describe the scatter between frames, not between recordings. The fluctuation of the active transduction, the variance $\kappa$ of Sec.~\ref{sec:active}, is measured from the frames themselves, as the excess of the in-phase over the quadrature variance of the frame amplitudes over the target.

The software used to automate experimental data acquisition was developed with assistance from Claude (Anthropic, Opus 5.5). The authors directed its development, reviewed and tested the code, and verified the results obtained using it.
\section{Readout noise temperature and operating point}\label{app:x0}

The receiver writes the RF voltage $v$ at its modulator onto the optical phase, $\phi=\pi v/V_\pi$. For balanced homodyne readout of a coherent carrier of power $P_{\rm opt}$ with detection efficiency $\eta$, shot noise gives the one-sided phase-noise power spectral density $h\nu/(2\eta P_{\rm opt})$, with $\nu$ the optical frequency~\cite{Caves1981}, that is, the voltage spectral density $(V_\pi/\pi)^2h\nu/(2\eta P_{\rm opt})$ at the modulator input (impedance $R$). A matched resistor at the temperature $T$ produces the one-sided voltage spectral density $k_BTR$ across the same input; equating the two gives the $T_{\rm q}$ of Sec.~\ref{sec:model}.

The RF field at the same port has the noise temperature $G_{\rm fe}T_A{}+T_N$, where $T_A$ is the antenna temperature (the sky brightness temperature averaged over the antenna pattern), $G_{\rm fe}$ the gain of any front end ($G_{\rm fe}=1$ for an antenna connected directly), and $T_N$ the front end's own noise at its output: $T_N=G_{\rm fe}(F-1)T_0$ for a front end of noise factor $F$, with $T_0=290$~K, and $T_N=0$ without one. Both noises are white across a channel, so $\sigma_k^2\propto T_{\rm q}$ and $P_k\propto G_{\rm fe}T_A{}+T_N$ with the same constant, and their ratio is the operating point $x$ of Sec.~\ref{sec:model}: the noise of the RF field in units of the coherent readout noise. The fractional rise of the noise floor with coherent readout is $x=N_{\rm on}/N_{\rm off}-1$, for the noise powers $N_{\rm on}$ and $N_{\rm off}$ with and without the RF signal; with squeezed readout the rise is $x/g$. The detector's electronic noise, more than 10~dB below the shot noise, is neglected here; in the experiment it is part of the measured readout noise and of $g$. Because the sky brightness falls as $f^{-2.55}$~\cite{Cane1979} while $T_{\rm q}$ is flat, $x$ changes by 15~dB across 5--20~MHz.

At HF the sky is bright: the median galactic noise gives $T_A=1.4\times10^{5}$~K at 12.5~MHz~\cite{ITU372}, the minimum at a radio-quiet site; man-made noise raises $T_A$, and $x_0$, further. Against this sky the noise of a front end referred to the antenna, $T_N/G_{\rm fe}=(F-1)T_0$, about 75~K for a noise figure of 1~dB, is negligible, and we set $T_N=0$ in the following. A link without a low-noise amplifier, with an integrated modulator ($V_\pi=3.5$~V, $\eta=0.9$, $\lambda=1550$~nm, $R=50~\Omega$), has $T_{\rm q}=6.4\times10^{6}$~K at $P_{\rm opt}=20~\upmu$W ($x_0=-16.7$~dB) and reaches $x_0=+0.3$~dB at 1~mW. The range $x_0=-20$ to $+5$~dB studied here corresponds to $P_{\rm opt}=9.3~\upmu$W to 2.9~mW on that link. At 1~mW, a 20-dB front-end amplifier raises $x_0$ to about $+20.3$~dB, where squeezing no longer matters. A replica of the scene applied to the modulator at a chosen RF power sets $x_0$ in the same way: it represents the antenna behind a noiseless gain $G_{\rm fe}$. Appendix~\ref{app:readout} gives $T_{\rm q}$ for the modulator of the experiment, measured in the records, and the range of $G_{\rm fe}$ that the replayed operating points correspond to. The sky model assumes that galactic noise reaches the antennas across the band, i.e., that the critical frequency of the ionospheric F2 layer lies below 5~MHz, as at night; otherwise the lowest channels are screened and $x$ falls there.

\section{Noise propagation in passive interferometric imaging}\label{app:passive}

This appendix derives how additive readout noise propagates from discrete-time samples to (i) the statistics of the sample visibility estimator and (ii) the variance of the synthesized, real-valued interferometric image, Eqs.~\eqref{eq:alphas} and~\eqref{eq:alphas_sq}, and how the PSNR gain depends on the operating point. The numerical examples of this appendix and of Appendix~\ref{app:active} use the design value $g=0.501$ (3~dB), the sky spectrum alone and a flat readout noise; the comparison with the measurements, at the measured $g$ and receiver spectra, is in Secs.~\ref{sec:passive_exp} and~\ref{sec:active_exp}.

\subsection{Statistical model}\label{app:passive_model}

In one frequency channel, receiver $k$ yields one complex sample $E_k[n]$ per block of its record, $n=1,\dots,N$ [Eq.~\eqref{eq:Ek}; Appendix~\ref{app:channels}]. $S_k[n]$ contains every RF contribution at the modulator: the emission correlated across the array and the sky emission uncorrelated between receivers. The fields $\{S_k[n]\}$ are jointly wide-sense-stationary (WSS), zero-mean, proper complex Gaussian processes. We model the channel samples of different blocks as independent and proper, an approximation justified by non-overlapping blocks much longer than the correlation time of the band-limited signals~\cite{Thompson2017}. The readout noise is zero-mean and independent across receivers and samples:
\begin{align}
  \E[\delta_k[n]] &= 0, \label{eq:c_noisemean}\\
  \E[\delta_k[n]\,\delta_{k'}^*[n']] &= \begin{cases}\sigma_k^2 & k=k',\ n=n'\\ 0 & \text{otherwise},\end{cases} \label{eq:c_noisevar}\\
  \E[S_k[n]\,\delta_{k'}^*[n']] &= 0. \label{eq:c_indep}
\end{align}
Different frequency channels are treated as independent in the same approximation; the broadband image is the sum of the channel images (Appendix~\ref{app:regimes}).

\subsection{Visibility statistics}\label{app:passive_vis}

For a baseline pair $(i,j)$, with the signal powers $P_i$, $P_j$ and noise variances $\sigma_i^2$, $\sigma_j^2$ of Sec.~\ref{sec:model}, the visibility between receivers $i\neq j$ at zero lag is
\begin{equation}
  V_{ij} \equiv \E\bigl[E_i[n]\,E_j^*[n]\bigr] = \E\bigl[S_i[n]\,S_j^*[n]\bigr], \label{eq:c_Vij}
\end{equation}
and, using independence and the zero-mean noise model, $\E|E_i[n]|^2=P_i+\sigma_i^2$. The estimator of Eq.~\eqref{eq:vis} is the mean of the products $E_i[n]E_j^*[n]$ over $N$ independent samples. By linearity of expectation and stationarity, $\E[\Vhat_{ij}]=\E[E_i[1]E_j^*[1]]=V_{ij}$: the estimator is unbiased. Because the products are i.i.d., $\Var(\Vhat_{ij})=\Var(E_iE_j^*)/N$ with $\Var(E_iE_j^*)=\E|E_iE_j^*|^2-|V_{ij}|^2$. For jointly Gaussian, zero-mean, proper complex variables, Isserlis' theorem~\cite{Isserlis1918} gives
\begin{equation}
\begin{aligned}
  \E\bigl[E_iE_j^*\,E_k^*E_{k'}\bigr] ={}& \E[E_iE_j^*]\,\E[E_k^*E_{k'}]\\
  &+ \E[E_iE_k^*]\,\E[E_{k'}E_j^*].
\end{aligned}
\label{eq:c_isserlis}
\end{equation}
With $k=i$, $k'=j$, it gives $\E\bigl[|E_i|^2|E_j|^2\bigr]=(P_i+\sigma_i^2)(P_j+\sigma_j^2)+|V_{ij}|^2$, hence
\begin{equation}
  \Var(\Vhat_{ij}) = \frac{1}{N}\bigl(P_iP_j + P_i\sigma_j^2 + P_j\sigma_i^2 + \sigma_i^2\sigma_j^2\bigr). \label{eq:c_varV}
\end{equation}
By the same theorem, the estimation errors of two baselines $(i,j)$ and $(k,k')$ have the covariance
\begin{equation}
\begin{aligned}
  \operatorname{Cov}(\Vhat_{ij},\Vhat_{kk'}) &\equiv \E\bigl[(\Vhat_{ij}-V_{ij})(\Vhat_{kk'}-V_{kk'})^*\bigr]\\
  &= \frac{1}{N}\,\E[E_iE_k^*]\,\E[E_{k'}E_j^*],
\end{aligned}
\label{eq:c_covV}
\end{equation}
which reduces to Eq.~\eqref{eq:c_varV} for $(k,k')=(i,j)$. It is non-zero for baselines that share a receiver and see correlated flux, e.g.\ $\operatorname{Cov}(\Vhat_{ij},\Vhat_{ik'})=(P_i+\sigma_i^2)V_{k'j}/N$, and even for disjoint baselines, $V_{ik}V_{k'j}/N$. These covariances vanish only when the scene has no correlated flux.

\subsection{Image variance}\label{app:passive_img}

With real synthesis weights $w_b$ and $\varphi_b(l,m)\equiv2\pi(u_bl+v_bm)$, Eq.~\eqref{eq:image} is, equivalently, a sum over ordered receiver pairs with $w_{ji}=w_{ij}$ and $\varphi_{ji}=-\varphi_{ij}$,
\begin{equation}
  \Ihat(l,m) = \sum_{i\neq j} w_{ij}\,\Vhat_{ij}\,e^{i\varphi_{ij}(l,m)}. \label{eq:c_pairs}
\end{equation}
Substituting $E_k=S_k+\delta_k$ into Eqs.~\eqref{eq:vis} and~\eqref{eq:c_pairs} splits the image into three parts,
\begin{equation}
  \Ihat = \Ihat_{SS} + \Ihat_{S\delta} + \Ihat_{\delta\delta}, \label{eq:c_split}
\end{equation}
formed from the products $S_iS_j^*$, $S_i\delta_j^*+\delta_iS_j^*$ and $\delta_i\delta_j^*$, respectively. $S$ and $\delta$ are independent and zero-mean Gaussian: $\Ihat_{SS}$ and $\Ihat_{\delta\delta}$ are functions of independent variables, and $\Ihat_{S\delta}$ is odd in both $S$ and $\delta$. The three parts are therefore uncorrelated, which gives Eq.~\eqref{eq:alphas} with
\begin{equation}
  \alpha_0\equiv\Var[\Ihat_{SS}],\quad \alpha_1\equiv\Var[\Ihat_{S\delta}],\quad \alpha_2\equiv\Var[\Ihat_{\delta\delta}]. \label{eq:c_alphadef}
\end{equation}
Each coefficient is a variance and hence non-negative. From Eqs.~\eqref{eq:c_covV} and~\eqref{eq:c_pairs}, with $\Gamma_{ik}\equiv\E[S_iS_k^*]$ ($\Gamma_{ii}=P_i$, $\Gamma_{ik}=V_{ik}$ for $i\neq k$),
\begin{align}
  \alpha_0 &= \frac{1}{N}\sum_{i\neq j}\sum_{k\neq k'} w_{ij}w_{kk'}\,e^{i(\varphi_{ij}-\varphi_{kk'})}\,\Gamma_{ik}\Gamma_{k'j}, \label{eq:c_a0}\\
  \alpha_1 &= \frac{1}{N}\sum_{i\neq j} w_{ij}\Bigl[\sigma_j^2\sum_{k\neq j}w_{kj}\,e^{i(\varphi_{ij}-\varphi_{kj})}\Gamma_{ik} \nonumber\\
  &\qquad + \sigma_i^2\sum_{k'\neq i}w_{ik'}\,e^{i(\varphi_{ij}-\varphi_{ik'})}\Gamma_{k'j}\Bigr], \label{eq:c_a1}\\
  \alpha_2 &= \frac{1}{N}\sum_{i\neq j} w_{ij}^2\,\sigma_i^2\sigma_j^2. \label{eq:c_a2}
\end{align}
Terms with $(k,k')\neq(i,j)$ are the covariances between different baselines of Eq.~\eqref{eq:c_covV}; they contain visibilities. When they are negligible (a weak or spatially resolved scene), only $(k,k')=(i,j)$ remains and, with $\sum_{i\neq j}=2\sum_b$,
\begin{equation}
\begin{aligned}
  \alpha_0 &= \frac{2}{N}\sum_b|w_b|^2P_iP_j,\\
  \alpha_1 &= \frac{2}{N}\sum_b|w_b|^2(P_i\sigma_j^2+P_j\sigma_i^2),\\
  \alpha_2 &= \frac{2}{N}\sum_b|w_b|^2\sigma_i^2\sigma_j^2,
\end{aligned}
\label{eq:c_aweak}
\end{equation}
the image-domain counterpart of Eq.~\eqref{eq:c_varV}. On bright point sources the cross-baseline terms of $\alpha_0$ dominate (self-noise, Appendix~\ref{app:regimes}).

\subsection{Squeezed readout and bounds}\label{app:passive_sq}

$\Ihat_{S\delta}$ is linear and $\Ihat_{\delta\delta}$ quadratic in the readout noise. Under $\sigma_k^2\to g\sigma_k^2$ for all $k$, $\alpha_1\to g\alpha_1$ and $\alpha_2\to g^2\alpha_2$, while $\alpha_0$ is unchanged, which is Eq.~\eqref{eq:alphas_sq}. With a frequency-dependent squeezing $g(f)$, the same substitution applies channel by channel, with $g(f)$ at the frequency $f$ of each channel, and the broadband coefficients are the sums of the channel coefficients (Appendix~\ref{app:regimes}); the predictions of Sec.~\ref{sec:passive_exp} with the measured $g(f)$ are evaluated in this way. Because $\alpha_0,\alpha_1,\alpha_2\ge0$ and $0<g<1$, the denominator of Eq.~\eqref{eq:gain_passive} lies between $g^2$ times the numerator and the numerator itself, which gives the bounds of Sec.~\ref{sec:passive}. The gain equals $G$ exactly where $\alpha_0=g\,\alpha_2$, independently of $\alpha_1$ [setting $g(\alpha_0+\alpha_1+\alpha_2)=\alpha_0+g\alpha_1+g^2\alpha_2$ gives $\alpha_0(1-g)=g\alpha_2(1-g)$].

\subsection{Dependence on the operating point}\label{app:regimes}

\subsubsection{Resolved sky}
For equal receivers ($P_k=\bar P$, $\sigma_k^2=\sigma^2$) at a pixel without correlated flux, Eq.~\eqref{eq:c_aweak} gives $\alpha_0:\alpha_1:\alpha_2=\bar P^2:2\bar P\sigma^2:\sigma^4$, so $\Var[\Ihat]\propto(\bar P+\sigma^2)^2$ and, with $x=\bar P/\sigma^2$, the single-channel gain of Sec.~\ref{sec:passive} follows. The signal--noise share of the variance, $2x/(x+1)^2$, never exceeds one half, so on resolved sky the gain lies between $G$ and $2G$ only for $x\le\sqrt g$.

\subsubsection{Point sources (self-noise)}
For a point source of power $P_{\rm pt}$ per receiver on top of uncorrelated power $P_u$, the cross-baseline terms of Eq.~\eqref{eq:c_a0} give, at the source pixel with unit weights~\cite{Kulkarni1989},
\begin{equation}
\begin{aligned}
  N\,\Var[\Ihat] ={}& M(M-1)\\
  &\times\Bigl\{\bigl[(M-1)P_{\rm pt}+P_u+\sigma^2\bigr]^2\\
  &\qquad+(M-1)P_{\rm pt}^2\Bigr\}.
\end{aligned}
\label{eq:c_kulkarni}
\end{equation}
The source power enters multiplied by the array gain $M-1$. The gain therefore follows the single-channel expression approximately, with $x$ replaced by $x_{\rm eff}=[(M-1)P_{\rm pt}+P_u]/\sigma^2$; because of the additional term $(M-1)P_{\rm pt}^2$, the substitution is not exact, but for $M=24$ and $g=0.501$ it reproduces Eq.~\eqref{eq:c_kulkarni} to within $0.04$~dB at every source power, the largest deviation occurring without uncorrelated power ($P_u=0$). Away from the source the self-noise variance equals $I_0^2(l,m)/N$, so the local gain is lowest where the noise-free image is bright.

\subsubsection{Full band}
With the channels treated as independent (Appendix~\ref{app:passive_model}), each $\alpha$ of the broadband image, the sum of the channel images, is the sum of the channel coefficients. With uniform channel weights the low channels, where $x$ is largest, carry most of the scene term $\alpha_0$ and the cross term $\alpha_1$ while $\alpha_2$ is the same in every channel, so the broadband gain at a given $x_0$ is lower than the single-channel expression evaluated at $x_0$ (Fig.~\ref{fig:validation}).

\subsubsection{Integration time}
Every $\alpha$ scales as $1/N$, inversely with the integration time $T_{\rm int}$, so a longer $T_{\rm int}$ improves both readouts equally and does not change the regime. Equal MSE needs integration times in the ratio $T_{\rm int}^{\rm coh}/T_{\rm int}^{\rm sq}=10^{\dPSNR/10}$, which on resolved sky is $[(x+1)/(x+g)]^2$.

\subsection{Monte Carlo simulation}\label{app:mc}

\begin{figure*}[t]
\centering
\includegraphics[width=0.8\textwidth]{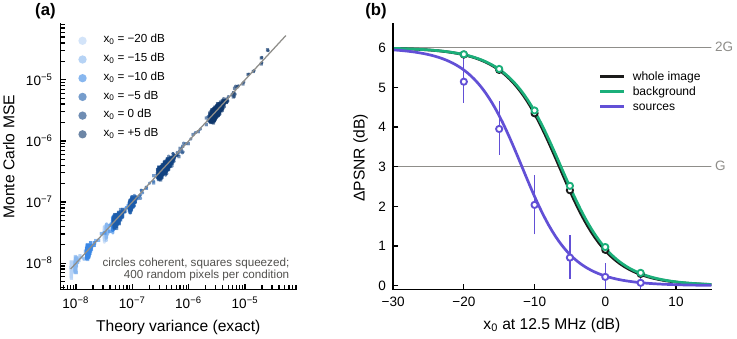}
\caption{Passive imaging: exact theory against Monte Carlo (208 independent skies, $g=0.501$). (a)~Pixel-wise MSE, Monte Carlo versus Eqs.~\eqref{eq:alphas} and~\eqref{eq:c_a0}--\eqref{eq:c_a2}; circles coherent, squares squeezed, 400 random pixels per condition. (b)~PSNR gain versus $x_0$ for the whole image, the sources ($I_0>50\%$ of the peak) and the background ($|I_0|<5\%$ of the peak): exact theory (lines) and Monte Carlo with 95\% bootstrap intervals (points); the same skies are used at every $x_0$, so the points are correlated. Gray lines: $G$ and $2G$.}
\label{fig:validation}
\end{figure*}

Figure~\ref{fig:validation} compares the exact image variance, Eqs.~\eqref{eq:alphas} and~\eqref{eq:c_a0}--\eqref{eq:c_a2}, with a Monte Carlo simulation of the complete passive chain for the parameters of Table~\ref{tab:scene_passive}: true-delay RF fields of 208 independent skies, readout noise with and without squeezing, FX correlation and multi-frequency imaging. Theory and simulation agree within the Monte Carlo scatter, and the field-averaged MSE agrees to within 1.1\% at every operating point. The simulation implements the same noise model, so the agreement verifies the derivation; the measurements of Sec.~\ref{sec:passive_exp} test the model. At $x_0=-10$~dB the simulation gives 4.42~dB (95\% interval $4.37$ to $4.46$) on the background (theory 4.40~dB) and 2.04~dB (95\% interval $1.30$ to $2.80$) on the sources (theory 2.25~dB). The background crosses $G$ at $x_0=-6.3$~dB and the sources at $-11.8$~dB, 5.5~dB lower, because their power is array-gained [Eq.~\eqref{eq:c_kulkarni}]; with the measured receiver spectra the crossings lie at the values given in Sec.~\ref{sec:passive_exp}.

\section{Noise propagation in active synthetic-aperture imaging}\label{app:active}

This appendix derives how additive readout noise propagates through the active synthetic-aperture imaging chain: matched filtering (range compression), fractional-delay interpolation, and coherent phase-compensated summation across pulses. It gives the terms of Eq.~\eqref{eq:bp}, the factor $\beta(q)$ of Eq.~\eqref{eq:varz}, the variance and MSE of the intensity image [Eqs.~\eqref{eq:gammas}--\eqref{eq:mse_active}] and the bounds of the PSNR gain [Eq.~\eqref{eq:gain_active}].

\subsection{Statistical model}\label{app:active_rc}

The receiver records the echo of each pulse directly in the 5--20~MHz band; $E_k[n]$ of Eq.~\eqref{eq:Ek} is its complex analytic representation, sampled at $f_s$ (at the times $t_n$), where $S_k[n]$ is the deterministic scene-dependent echo for a fixed waveform, platform trajectory and scene reflectivity. The replica $c(t)$ of Eq.~\eqref{eq:rc} is the analytic representation of the transmitted chirp, and $f_c=12.5$~MHz the center of its band. Consistent with Appendix~\ref{app:passive_model}, $\delta_k[n]$ is zero-mean, WSS, proper complex Gaussian and independent across aperture positions. The matched filter passes only the signal band, so only the in-band noise level enters the results. We therefore model $\delta_k[n]$ as white, with the variance $\sigma_k^2$ per sample [Eqs.~\eqref{eq:c_noisemean} and~\eqref{eq:c_noisevar}], at the spectral density of the readout noise in the band. The records themselves are band-limited to $B=15$~MHz: their samples are correlated, and their noise power is $(B/f_s)\,\sigma_k^2$; for active imaging, Eq.~\eqref{eq:coh} and Sec.~\ref{sec:active} use the variance $\sigma_k^2$ of this white model, and ratios of noise powers, such as $g$, are the same for both. Because the noise is proper, its pseudo-covariance vanishes as well,
\begin{equation}
  \E\bigl[\delta_k[n]\,\delta_{k'}[n']\bigr] = 0.
  \label{eq:d_proper}
\end{equation}

\subsection{Focused-field statistics}\label{app:active_bp}

Inserting Eq.~\eqref{eq:rc} into Eq.~\eqref{eq:bp}, the two phase factors cancel, $e^{i\varphi_k(q)}e^{-i2\pi f_c\tau_{k,q}}=1$, and the focused field is a fixed linear combination of the samples of all $P$ records,
\begin{equation}
  z_q = \sum_kw_k(q)\sum_n E_k[n]\,c^*(t_n-\tau_{k,q}).
  \label{eq:d_zlin}
\end{equation}
With $E_k[n]=S_k[n]+\delta_k[n]$, the two terms of Eq.~\eqref{eq:bp} are the deterministic signal field and the noise
\begin{equation}
\begin{aligned}
  z_{0,q} &= \sum_kw_k(q){}\sum_n S_k[n]\,c^*(t_n-\tau_{k,q}),\\
  n_q &= \sum_kw_k(q){}\sum_n \delta_k[n]\,c^*(t_n-\tau_{k,q}).
\end{aligned}
\label{eq:d_z0n}
\end{equation}
Since $n_q$ has zero mean [Eq.~\eqref{eq:c_noisemean}], $\E[z_q]=z_{0,q}$, so the focused field is unbiased, and $\Var(z_q)=\Var(n_q){}=\E|n_q|^2$. Because the noise is independent across aperture positions and samples [Eq.~\eqref{eq:c_noisevar}], the cross terms of $\E|n_q|^2$ with $k\neq k'$ or $n\neq n'$ vanish, and
\begin{equation}
\begin{aligned}
  \Var(z_q) ={}& \sum_{k,k'}w_k(q)\,w_{k'}^*(q)\sum_{n,n'}\E\bigl[\delta_k[n]\,\delta_{k'}^*[n']\bigr]\\
  &{}\times c^*(t_n-\tau_{k,q})\,c(t_{n'}-\tau_{k',q})\\
  ={}& \sum_k\sigma_k^2|w_k(q)|^2{}\sum_n|c(t_n-\tau_{k,q})|^2.
\end{aligned}
\label{eq:d_varz}
\end{equation}
The inner sum is the energy of the replica, $E_c\equiv\sum_n|c(t_n-\tau)|^2$. It is the same for every pulse and pixel, because the receive window (Table~\ref{tab:scene_active}) contains the whole replica at every delay of the image. The same receiver is used for all aperture positions, so $\sigma_k^2\equiv\sigma^2$ is uniform across $k$ and factors out of the sum:
\begin{equation}
  \sigma_z^2(q) \equiv \Var(z_q) {}= \sigma^2E_c\sum_{k=1}^{P}|w_k(q)|^2, \label{eq:d_beta}
\end{equation}
which is Eq.~\eqref{eq:varz} with $\beta(q)=E_c\sum_{k=1}^{P}|w_k(q)|^2>0$. The same expansion with Eq.~\eqref{eq:d_proper} gives $\E[n_q^2]=0$: as a sum of proper complex Gaussian samples with fixed coefficients, $n_q$ is itself proper complex Gaussian, $n_q\sim\mathcal{CN}(0,\sigma_z^2(q))$.

Equation~\eqref{eq:rc} defines $r_k(\tau)$ at any delay, because the replica is a known function. In the processing, $r_k(\tau)$ is computed at delays spaced by $\Delta\tau=2.5$~ns (the output of the matched filter, restricted to the 5--20~MHz band and upsampled by 4), and its value at $\tau_{k,q}$ is interpolated linearly. With $\tau'$ the grid delay just below $\tau_{k,q}$ and $\theta_{k,q}\equiv(\tau_{k,q}-\tau')/\Delta\tau\in[0,1)$,
\begin{equation}
  r_k(\tau_{k,q}) \simeq (1-\theta_{k,q})\,r_k(\tau')+\theta_{k,q}\,r_k(\tau'+\Delta\tau).
  \label{eq:d_interp}
\end{equation}
By Eqs.~\eqref{eq:rc} and~\eqref{eq:c_noisevar}, the noise parts of $r_k(\tau)$ and $r_k(\tau+\Delta)$ have the correlation $\sigma_k^2R(\Delta)$, where $R(\Delta)\equiv e^{i2\pi f_c\Delta}\sum_nc^*(t_n-\tau)\,c(t_n-\tau-\Delta)$ is the autocorrelation of the replica at baseband; like $E_c$, it does not depend on $\tau$, and $R(0)=E_c$. The interpolation therefore replaces the inner sum $E_c$ of Eq.~\eqref{eq:d_varz} by
\begin{equation}
  \bigl[(1-\theta_{k,q})^2+\theta_{k,q}^2\bigr]E_c+2\theta_{k,q}(1-\theta_{k,q})\,\mathrm{Re}\,R(\Delta\tau).
  \label{eq:d_linint}
\end{equation}
The range-compressed output is oversampled: its main lobe, of width $\approx1/B$, spans many grid steps, and with the RF carrier removed its phase advances by at most $2\pi\times7.5~\mathrm{MHz}{}\times \Delta\tau=0.12$~rad per step. Hence $\mathrm{Re}\,R(\Delta\tau)\ge0.99\,E_c$. Computed from the chirp spectrum, the matched filter and the upsampling of the experiment, $\mathrm{Re}\,R(\Delta\tau)=0.998\,E_c$, so Eq.~\eqref{eq:d_linint} equals $E_c$ to within 0.1\%. The band limit of the matched filter lowers the variance by another 0.16\%, so that in the processing $\sigma_z^2(q)$ is lower than Eq.~\eqref{eq:d_beta} by less than 0.3\%. The predictions of Sec.~\ref{sec:active_exp} take $\sigma_z^2(q)$ directly from the variance of the measured coherent images over the frames, which does not assume a flat noise spectrum.

\subsection{Image variance}\label{app:active_img}

The active image is the intensity $I_q\equiv|z_q|^2$. Substituting $z_q=z_{0,q}+n_q$, with $n_q\sim\mathcal{CN}(0,\sigma_z^2(q))$, splits it into three parts,
\begin{equation}
  I_q = I_{SS}+I_{S\delta}+I_{\delta\delta},
  \label{eq:d_split}
\end{equation}
with $I_{SS}=|z_{0,q}|^2$, $I_{S\delta}=2\,\mathrm{Re}(z_{0,q}^*n_q)$ and $I_{\delta\delta}=|n_q|^2$, the counterparts of the three parts of Eq.~\eqref{eq:c_split}. $I_{SS}$ is deterministic: unlike $\Ihat_{SS}$, it does not fluctuate, so there is no counterpart of $\alpha_0$. $I_{S\delta}$ is odd and $I_{\delta\delta}$ even in $n_q$, and the odd moments of the circular noise vanish; the two are therefore uncorrelated, which gives Eq.~\eqref{eq:gammas} with
\begin{equation}
  \gamma_1\equiv\Var[I_{S\delta}],\qquad \gamma_2\equiv\Var[I_{\delta\delta}].
  \label{eq:d_gammadef}
\end{equation}
Because $n_q$ is proper, $\E[n_q^2]=0$, so $I_{S\delta}=z_{0,q}^*n_q+z_{0,q}n_q^*$ has zero mean and the variance $2|z_{0,q}|^2\,\E|n_q|^2$. Isserlis' theorem [Eq.~\eqref{eq:c_isserlis}], with all four variables equal to $n_q$, gives $\E|n_q|^4=2\sigma_z^4(q)$, so $I_{\delta\delta}$ has the mean $\sigma_z^2(q)$ and the variance $\sigma_z^4(q)$. Hence
\begin{equation}
  \gamma_1 = 2|z_{0,q}|^2\sigma_z^2(q),\qquad \gamma_2 = \sigma_z^4(q),
  \label{eq:d_gammas}
\end{equation}
which with $\sigma_z^2(q)=\beta(q)\sigma^2$ are the coefficients of Eq.~\eqref{eq:gammas}. The mean of $I_{\delta\delta}$ biases the intensity, $\E[I_q]=|z_{0,q}|^2+\sigma_z^2(q)$, and the squared bias, $\sigma_z^4(q)=\gamma_2$, adds to the variance in the MSE, which gives Eq.~\eqref{eq:mse_active}.

\subsection{Squeezed readout and bounds}\label{app:active_sq}

$I_{S\delta}$ is linear and $I_{\delta\delta}$ quadratic in the readout noise. Under $\sigma^2\to g\sigma^2$, Eq.~\eqref{eq:d_beta} gives $\sigma_z^2(q)\to g\,\sigma_z^2(q)$, so $\gamma_1\to g\gamma_1$ and $\gamma_2\to g^2\gamma_2$, which is Eq.~\eqref{eq:gammas_sq}; the squared bias, equal to $\gamma_2$, scales in the same way, which gives $\mathrm{MSE}_{\rm sq}(I_q)$ in Eq.~\eqref{eq:mse_active}. Because $\gamma_1,\gamma_2\ge0$ and $0<g<1$, the denominator of Eq.~\eqref{eq:gain_active} lies between $g^2$ and $g$ times the numerator, which gives the bounds of Sec.~\ref{sec:active}: the gain lies between $G$ and $2G$. The lower bound is $G$, and not zero as in passive imaging, because the deterministic echo has no scene term. For a single pixel, $\gamma_1/\gamma_2=2|z_{0,q}|^2/\sigma_z^2(q)=2\rho_q$, with the local SNR $\rho_q\equiv|z_{0,q}|^2/\Var(z_q)$, and Eq.~\eqref{eq:gain_active} becomes
\begin{equation}
\begin{aligned}
  \dPSNR &{}= 10\log_{10}\frac{\rho_q+1}{g\,(\rho_q+g)}\\
  &{}= G+10\log_{10}\frac{\rho_q+1}{\rho_q+g},
\end{aligned}
\label{eq:d_gain1}
\end{equation}
which decreases from $2G$ at $\rho_q\ll g$ to $G$ at $\rho_q\gg1$.

These bounds hold when the readout noise is the only noise (for classical noise in the echoes, see Appendix~\ref{app:active_ext}). A fluctuation of the RF-to-optical transduction from frame to frame multiplies the echoes of a frame by a real factor of mean 1 and variance $\kappa$. To first order in $\kappa$, it multiplies $I_{SS}$ by a factor of mean $1+\kappa$ and variance $4\kappa$, which adds $4\kappa|z_{0,q}|^4$ to the variance and to the MSE of either readout; through the readout noise, it also adds $4\kappa|z_{0,q}|^2\sigma_z^2(q)$ to the MSE, smaller by the factor $\sigma_z^2(q)/|z_{0,q}|^2$. Squeezing does not reduce the first term, which therefore lowers the gain below $G$ where $|z_{0,q}|^2$ is large.

\subsection{Extensions and Monte Carlo simulation}\label{app:active_ext}

\subsubsection{Distributed targets}
If the reflectivity of a resolution cell is itself random (fully developed speckle, $z_{0,q}\sim\mathcal{CN}(0,\mu_q)$), the image fluctuates even without readout noise, the active counterpart of $\alpha_0$. Against the mean intensity $\mu_q$, $\mathrm{MSE}=(\mu_q+\sigma_z^2)^2+\sigma_z^4$~\cite{Goodman2007}: the gain falls to zero at high SNR and equals $G$ at $\mu_q/\sigma_z^2=\sqrt{2g}=1.00$. This MSE refers the image to the ensemble-mean intensity, which a speckle image estimates; for one fixed realization of the reflectivity, referred to its own $|z_{0,q}|^2$, the deterministic results of Sec.~\ref{sec:active} apply, with a gain between $G$ and $2G$.

\subsubsection{Classical noise in the echoes}
If the echoes also carry classical noise (sky noise, front end) with operating point $x$ (Sec.~\ref{sec:model}), the additive noise of the samples has the variance $(x+1)\sigma^2$ with coherent and $(x+g)\sigma^2$ with squeezed readout: only the readout part is reduced, and every result of Sec.~\ref{sec:active} and of this appendix holds with $\sigma^2$ replaced by $(x+1)\sigma^2$ and $g$ replaced by $(x+g)/(x+1)$. For $x\ll g$ the full squeezing level applies.

\subsubsection{Frame averaging}
Averaging the complex images $z_q$ of $K$ independent frames coherently, before the magnitude or the intensity is formed, divides $\sigma_z^2$ by $K$, so equal image quality needs $K_{\rm sq}=g\,K_{\rm coh}$ frames, for any image metric and SNR (Sec.~\ref{sec:discussion}). Averaging the intensity images instead reduces only their variance, $\mathrm{MSE}(\bar I_q)=(\gamma_1+\gamma_2)/K+\gamma_2$, because the bias $\sigma_z^2$ remains, and does not scale in this way.

\subsubsection{Monte Carlo simulation}
For the active model of Table~\ref{tab:scene_active}, the intensity gain on the target is 5.99~dB at an echo SNR of $-50$~dB and 3.01~dB at 0~dB, never below $G$; a Monte Carlo simulation with 200 noise realizations agrees with Eq.~\eqref{eq:gain_active} within its 95\% intervals at every SNR.

\end{document}